\documentclass[iop,apj,tighten,numberedappendix,twocolappendix]{emulateapj}

\usepackage{apjfonts}
\usepackage{amsmath}

\newcommand{\er}{$L_{\rm bol}/L_{\rm Edd}$}
\newcommand{\fvar}{$F_{\rm var}$}
\newcommand{\rmax}{$r_{\rm max}$}
\newcommand{\rl}{$R_{\rm BLR}$--$L$}
\newcommand{\ks}{$\chi^2$}
\newcommand{\kms}{$\rm km~s^{-1}$}
\newcommand{\rfe}{$R_{\rm Fe}$}

\newcommand{\thb}{$\tau_{\rm H\beta}$}
\newcommand{\tfe}{$\tau_{\rm Fe}$}

\newcommand{\fagn}{$F_{\rm AGN}$}
\newcommand{\ffe}{$F_{\rm Fe}$}
\newcommand{\fgal}{$F_{\rm gal}$}
\newcommand{\fhb}{$F_{\rm H\beta}$}
\newcommand{\fsumc}{$F_{\rm 5100}$}

\newcommand{\fwhmhb}{${\rm FWHM}_{\rm H\beta}$}
\newcommand{\fwhmfe}{${\rm FWHM}_{\rm Fe}$}

\newcommand{\vfe}{$V_{\rm Fe}$}

\newcommand{\cafv}{[Ca {\sc v}]}
\newcommand{\feii}{Fe {\sc ii}}

\newcommand{\fefvi}{[Fe {\sc vi}]}
\newcommand{\fefvii}{[Fe {\sc vii}]}

\newcommand{\hb}{H$\beta$}
\newcommand{\hc}{H$\gamma$}
\newcommand{\hei}{He {\sc i}}
\newcommand{\heii}{He {\sc ii}}
\newcommand{\nai}{Na {\sc i}}
\newcommand{\nfi}{[N {\sc i}]}
\newcommand{\oiii}{[O {\sc iii}]}

\newcommand{\iras}{IRAS 04416$+$1215}
\newcommand{\mcg}{MCG $+$06$-$26$-$012}
\newcommand{\irasf}{IRAS F12397$+$3333}

\slugcomment{Accepted for publication in \textit{The Astrophysical Journal}}

\shorttitle{\feii\ Reverberation in Nine Seyferts}
\shortauthors{Hu et al.}

\begin{document}

\title{Supermassive Black Holes with High Accretion Rates in Active Galactic
Nuclei. \\III. Detection of \feii\ Reverberation in Nine Narrow-Line Seyfert 1
Galaxies}

\author{Chen Hu\altaffilmark{1},
Pu Du\altaffilmark{1},
Kai-Xing Lu\altaffilmark{2,1},
Yan-Rong Li\altaffilmark{1},
Fang Wang\altaffilmark{6},
Jie Qiu\altaffilmark{1},
Jin-Ming Bai\altaffilmark{6},\\
Shai Kaspi\altaffilmark{7},
Luis C. Ho\altaffilmark{4,5},
Hagai Netzer\altaffilmark{7},
Jian-Min Wang\altaffilmark{1,3,*}\\
(SEAMBH collaboration)}

\altaffiltext{1}{Key Laboratory for Particle Astrophysics, Institute of High
Energy Physics, Chinese Academy of Sciences, 19B Yuquan Road, Beijing 100049,
China}

\altaffiltext{2}{Department of Astronomy, Beijing Normal University, Beijing
100875, China}

\altaffiltext{3}{National Astronomical Observatories of China, Chinese Academy
of Sciences, 20A Datun Road, Beijing 100020, China}

\altaffiltext{4}{Kavli Institute for Astronomy and Astrophysics, Peking
University, Beijing 100871, China} 

\altaffiltext{5}{Department of Astronomy, School of Physics, Peking
University, Beijing 100871, China} 

\altaffiltext{6}{Yunnan Observatories, Chinese Academy of Sciences, Kunming
650011, China}

\altaffiltext{7}{School of Physics and Astronomy and the Wise Observatory,
Tel-Aviv University, Tel-Aviv 69978, Israel}

\altaffiltext{*}{Corresponding author}

\begin{abstract}
  This is the third in a series of papers reporting on a large
  reverberation-mapping campaign aimed to study the properties of active
  galactic nuclei (AGNs) with high accretion rates. We present new results on
  the variability of the optical \feii\ emission lines in 10 AGNs observed by
  the Yunnan Observatory 2.4m telescope during 2012--2013. We detect
  statistically significant time lags, relative to the AGN continuum, in nine
  of the sources. This accurate measurement is achieved by using a
  sophisticated spectral fitting scheme that allows for apparent flux
  variations of the host galaxy, and several narrow lines, due to the changing
  observing conditions. Six of the newly detected lags are indistinguishable
  from the \hb\ lags measured in the same sources. Two are significantly
  longer and one is slightly shorter. Combining with \feii\ lags reported in
  previous studies, we find a \feii\ radius--luminosity relationship similar
  to the one for \hb, although our sample by itself shows no clear
  correlation. The results support the idea that \feii\ emission lines
  originate in photoionized gas which, for the majority of the newly reported
  objects, is indistinguishable from the \hb-emitting gas. We also present a
  tentative correlation between the lag and intensity of \feii\ and \hb\ and
  comment on its possible origin.
\end{abstract}

\keywords{galaxies: active --- galaxies: nuclei --- galaxies: Seyfert --- 
methods: data analysis --- quasars: emission lines}

\section{Introduction}

Most active galactic nuclei (AGNs) show prominent \feii\ emission lines in
their spectra. The lines appear in several broad bands that represent
thousands of individual transitions. The stronger bands cover the ranges of
4000--5400 \AA\ (hereafter optical \feii\ lines), 2800--3500 \AA\ and
2000--2600 \AA\ (hereafter UV \feii\ lines; e.g.,
\citealt{wills85,sulentic00,hu08b} and references therein). The strongest
\feii\ lines (relative to \hb) are observed in narrow-line Seyfert 1 galaxies
(NLS1s, see, e.g., \citealt{osterbrock85,boller96}). Such objects are
characterized by: 1) narrow ($<$ 2000 \kms) broad emission lines; 2) weak
\oiii\ line; 3) and steep hard X-ray spectrum. These objects are found usually
at the extreme end of the so-called Eigenvector 1 sequence
\citep{boroson92,sulentic00,shen14} indicating high Eddington ratios and
several other properties that are not fully understood.

Some NLS1s show evidence for super-Eddington accretion, with \er\ $>$ 1. We
refer to these objects as super-Eddington accreting massive black holes
(SEAMBHs). Our earlier work \citep{wang13,wang14} show that such objects are
potentially a new kind of standard candle for cosmology. To test this idea,
and to study in more detail the physical properties of extreme NLS1s, we have
initiated a large reverberation mapping (RM) campaign to measure, accurately,
the black hole (BH) mass of such sources. The initial results of this
campaign have been published in \citet[hereafter Paper I]{du14} and
\citet[Paper II]{wang14}. The present paper, the third in this series, is
dedicated to the study of \feii\ emission lines in our sample of SEAMBH
candidates. A fourth paper, submitted in parallel to this one, presents new
data on \hb\ time lags and BH mass in five extreme SEAMBHs and focuses on the
modification of the \rl\ relationship in AGNs
\citep[e.g.,][]{kaspi00,kaspi05,bentz13} in the presence of such sources.

The excitation mechanism of the \feii\ emission in AGNs has been discussed in
numerous publications
\citep[e.g.,][]{collin80,netzer83,joly87,sigut98,baldwin04,ferland09}. Most of
these studies suggest an origin in the broad-line region (BLR) but the line
intensities calculated so far are in poor agreement with most observations.
Other suggestions connect these lines to the outer part of the central
accretion disk \citep[e.g.,][]{joly87} but the agreement with observations is
still poor. In particular, none of the existing models can explain,
satisfactorily, the relative intensity of the UV and optical \feii\ lines and
the observed \feii\ spectrum in the range 2000--2600 \AA. Line variability is
an important tool in such studies since it can indicate, given a measured time
lag relative to the continuum variations, the location of the \feii-emitting
gas. Such variations have been detected in a number of sources (e.g.,
\citealt{boksenberg77,maoz93,kollatschny00,vestergaard05,wang05,kuehn08,shapovalova12}),
but robust lag measurements were not obtained. This situation has changed,
recently. \citet{bian10} revisited the data of PG 1700$+$518 from
\citet{kaspi00} and measured the light curve of the optical \feii\ lines. They
obtained a significant time lag relative to the 5100 \AA\ continuum albeit
with a very large uncertainty. \citet{rafter13} and \citet{chelouche14}
adopted the multivariate correlation function (MCF) scheme of
\citet{chelouche13} designed for photometric RM. \citet{rafter13} studied the
NLS1 SDSS\,J113913.91+335551.1 using RM and found the \feii\ time lag is
consistent with the one of the \hb\ line. \citet{chelouche13} measured optical
\feii\ light curves, and time lags, for three of the objects in
\citet{kaspi00}, one of which is the object studies by \citet{bian10}. They
also find somewhat less significant \feii\ time lag for three other sources.
\citet{chelouche13} presents a tentative \feii\ size-luminosity relation and
suggest that the \feii\ emission-region size is comparable to that of the \hb\
line. \citet{barth13} used spectroscopic measurements of optical \feii\ lines
and found lags that are 1.5 and 1.9 times longer than the corresponding \hb.
All these studies confirm the photoionization origin of the \feii\ lines. The
success of the \citet{barth13} campaign is due both to the detailed and
frequent spectroscopic observations and a novel method they used to correct
for the host galaxy contribution to the optical spectra.

This paper presents the results obtained, so far, to measure the time lags of
the optical \feii\ lines (hereafter \feii\ lines) in our SEAMBH campaign. We
have attempted to detect such lags in 10 of the sources and were able to
obtain statistically significant results in nine of them. In six of the newly
measured sources, the \feii\ lag is entirely consistent with the \hb\ lag and
in two others it is considerably longer. Section \ref{sec-obs} gives a brief
review of the observations and data reduction. Section \ref{sec-method}
describes our spectral fitting method, with emphasis on host galaxy and narrow
line subtraction which we find to be crucial to the analysis. More details are
given in Appendix \ref{sec-host}. Section \ref{sec-result} presents \feii\
light curves and their analysis for both \feii\ and \hb, and compares the
results of the new \hb\ lags to those presented in Paper I and Paper II. In
Section \ref{sec-discussion}, we plot the size-luminosity relation for \feii\
and compare the lag and intensity of \feii\ with \hb. The implications and
some additional interpretation are also discussed. Section \ref{sec-summary}
gives a summary of the new results.

\section{Observations and Data Reductions}
\label{sec-obs}

The details of the SEAMBH campaign, including the observations, data reduction
and analysis, were presented in Paper I and Paper II. For completeness, we
summarize the more important points below and discuss in detail the new method
of galaxy and narrow line subtraction.

\subsection{Sample}

Ten NLS1s identified as SEAMBH candidates were observed, spectroscopically and
photometrically, between October 2012 and June 2013. Objects names and
coordinates are listed in Table 1 of Paper II. \hb\ time lags for three of the
sources (Mrk 335, Mrk 142, \irasf) are presented in Paper I and for five
additional sources (Mrk 1044, Mrk 382, \mcg, Mrk 486, Mrk 493) in Paper II.
Like many other NLS1s \citep[e.g.]{boroson92,sulentic00,boroson02,zhou06}, all
the sources in our sample show strong \feii\ emission lines and high \er. This
sample is, therefore, different from most AGNs in the local universe and none
of the results presented below should be compared with earlier studies, like
\cite{hu08b} that address the general population properties.

\subsection{Spectroscopy and Data Reduction}
\label{sec-reduction}

The spectra were obtained using the Yunnan Faint Object Spectrograph and
Camera (YFOSC), mounted on the Lijiang 2.4m telescope at Yunnan Observatory of
the Chinese Academy of Sciences. A longslit with projected width of
2$\farcs$5 was oriented to take the spectra of the object and a nearby
non-varying comparison star simultaneously following, e.g., \citet{maoz90} and
\citet{kaspi00}. The comparison star is then used as a standard for flux
calibration. For the Lijiang 2.4m telescope, the rotator is accurate and the
tracking is stable. Thus the object and the comparison star were kept  within
the slit during the typical 30 min. exposures. The distance between the object
and the comparison star along the direction of slit width keeps less than 1
pixel (0$\farcs$283).

Grism 14 was used and yielded spectra covering the wavelength range of
3800--7200 \AA\ with a dispersion of 1.8 \AA\ pixel$^{-1}$. The final spectral
resolution, obtained by comparing the width of the \oiii\ emission line with
the one measured from the Sloan Digital Sky Survey (SDSS; \citealt{york00})
spectrum of the same object, is roughly 500 \kms. All spectra are extracted in
a uniform, large aperture of 8$\farcs$5 to minimize light losses. The flux
calibration is based on the comparison of the object flux to the comparison
star flux in the 2$\farcs$5$\times$8$\farcs$5 apertures. However, the host
galaxy flux calibration is different because of the non-stellar image --- the
galaxy is extended and resolved. This can result in apparent flux variations
due to variable seeing and mis-centering. This requires a different
calibration procedure that has a large impact on the various light curves.
Appendix \ref{sec-host} gives details of this procedure.

\section{Line and Continuum Measurements}
\label{sec-method}

\subsection{Fitting scheme}

The traditional method to measure the flux of the continuum and the broad
emission lines, in most RM studies, is simple integration
\citep[e.g.,][]{kaspi00}. A straight line is set between two line-free windows
to define the AGN continuum, and the flux of the emission line is measured by
simple integration above the line. This method works well for single, strong
emission lines, e.g., \hb, because there are no (or only weak) other emission
lines in this range and the wavelength window is narrow enough for the
continuum to be approximated by a straight line. In the case of \feii, neither
condition is satisfied.

\feii\ emission consists of thousands of lines that form a pseudo-continuum.
The most prominent features in the optical band are two bumps between 4500 to
5500 \AA: one between \hb\ and \hc, and the other on the red side of \oiii\
$\lambda$5007. Even for quasars, whose host galaxies are extremely faint
relative to the AGN, it is hard to find ``pure'' continuum windows, and it is
inappropriate to assume a straight line over such a wide wavelength range of
more than 1000 \AA. The strong host galaxy contribution in most of our
objects, and the contamination by other emission lines (e.g., \heii\ and
coronal lines), make the situation much worse. Appendix \ref{sec-feintg}
presents the \feii\ light curves measured by the traditional integration
method. All light curves have large scatter, and only less than a half show
rough structures. A simultaneous fitting including the \feii\ emission and all
the other spectral components is necessary. 

Template-fitting is a widely used method to measure \feii\ emission in
single-epoch spectra of AGNs (see, e.g., \citealt{hu08b} for a brief
overview). Input-output simulations show that template-fitting is a reliable
measurement of \feii\ emission with equivalent width (EW) $>$ 25 \AA\ in
quasars \citep{hu08b}. It is also common to include a host galaxy component
into the fitting of Type I AGNs with strong host contribution
\citep[e.g.,][]{zhou06,ho09}. The method has been adopted, recently, to
measure the light curves in a few reverberation mapping studies, and proved to
be successful. \citet{bian10} reanalyzed the spectra of PG 1700$+$518, using
data from \citet{kaspi00}. They detected a lag of $209^{+100}_{-147}$ days for
\feii\ emission. \citet{barth13} performed a very careful spectral
decomposition including a power-law continuum, a host galaxy component, \feii\
template, and other emission lines (\hb, \oiii, \heii, and \hei). Using this
technique they were able to obtain statistically significant reverberation
lags of the \feii\ lines in two Seyfert 1 galaxies, NGC 4593 and Mrk 1511.

The spectral fitting in the present paper follows the algorithm of
\citet{hu12}, where several spectral components are fitted simultaneously by
minimizing the \ks\ via Levenberg-Marquardt method. All the light curves are
obtained directly from the results of the fitting. Before fitting, we correct
the calibrated spectra for Galactic extinction and redshift. We use the
$R_V$-dependent Galactic extinction law given by \citet{cardelli89} and
\citet{odonnell94} and $R_V$ is assumed to be 3.1. The redshift $z$ and
$V$-band extinction are taken from the NASA/IPAC Extragalactic Database%
\footnote{
\url{http://ned.ipac.caltech.edu/}
}
and \citet{schlafly11}, as listed in Papers I and II.

\begin{figure*}
  \centering
  \includegraphics[angle=-90,width=0.45\textwidth]{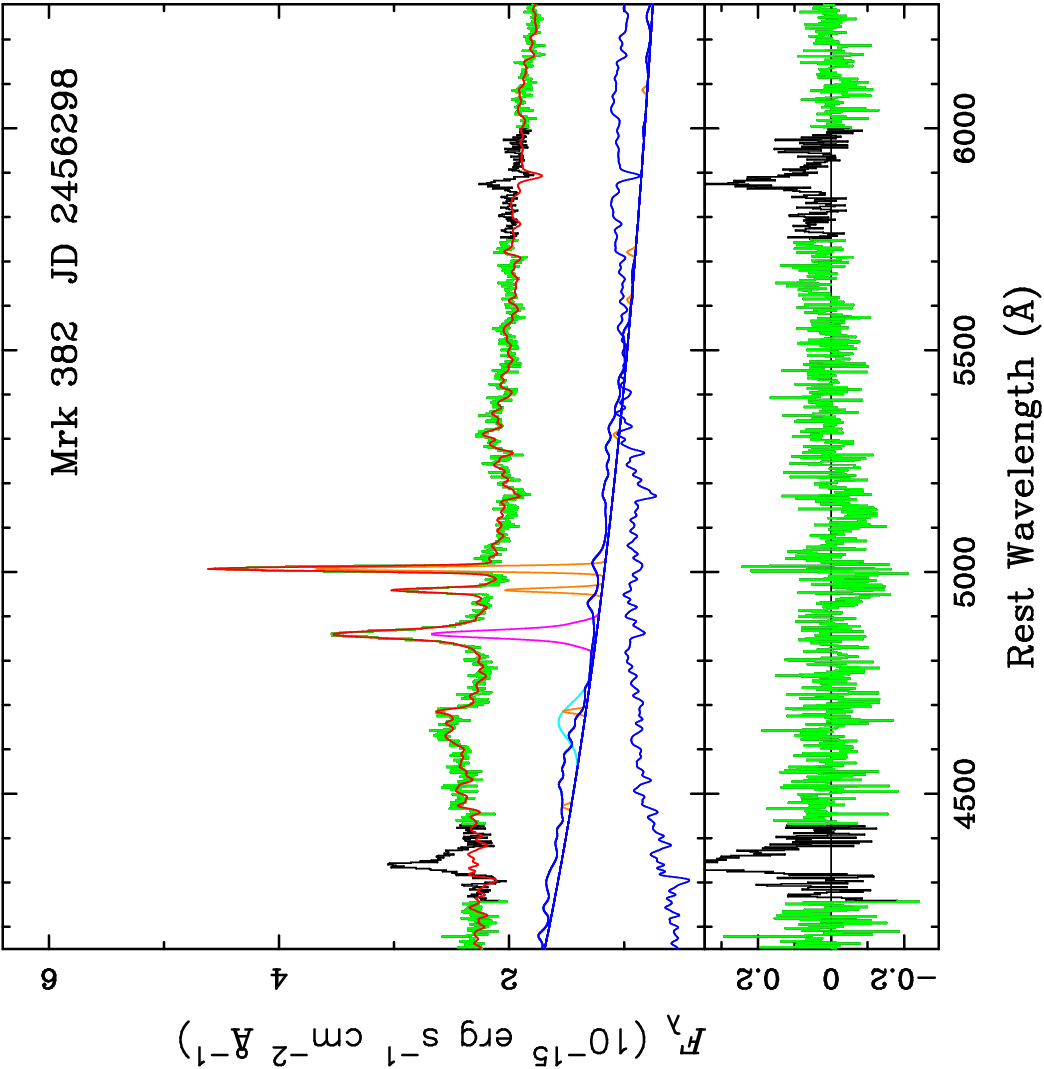}
  \hspace{0.05\textwidth}
  \includegraphics[angle=-90,width=0.45\textwidth]{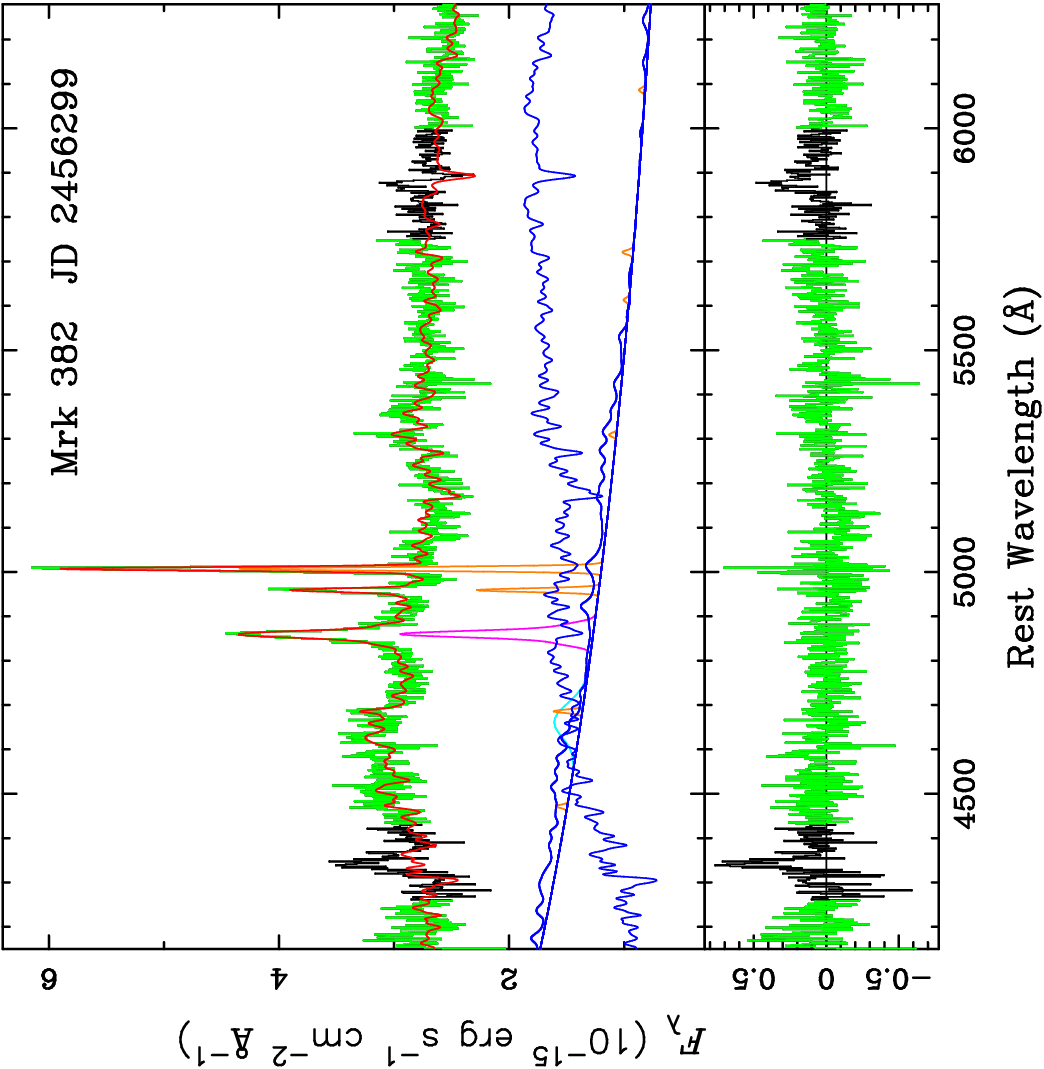}
  \caption{
  Individual-night spectra and multi-component model fitting for Mrk 382 taken
  at JDs 2456298 (left) and 2456299 (right). For each night, the top panel
  shows the Galactic extinction and redshift corrected spectrum (green or
  black for the parts in/out of the fitting window), and best-fit model (red).
  The model includes a pseudo-continuum (blue, composed by AGN power-law
  continuum, \feii\ emission, and host galaxy) and several emission lines
  (\hb\ in magenta, broad \heii\ $\lambda$4686 in cyan, and narrow emission
  lines in orange). The bottom panel shows the residuals. Note the large
  change in the host galaxy intensities for the two successive nights, which
  is not real but caused by our method of observation and calibration (see
  Appendix \ref{sec-host} for details).
  }
  \label{fig-spec382}
\end{figure*}

The left panel of Figure \ref{fig-spec382} shows the fitting to the spectrum
of Mrk 382 taken at Julian date (JD) 2456298, after Galactic extinction and
redshift correction. The fitting is performed in the rest frame wavelength
range 4150--6280 \AA, except two narrow windows around \hc\ and \hei\
$\lambda$5876. The two lines are not blended with the major part of \feii\
emission, and their study is beyond the scope of the present paper. We keep
them out of the fitting to avoid introducing too many unnecessary parameters.
The observed spectra in the fitting windows are plotted in green, while those
left out of the fitting are in black. The fitting includes the following
components: (1) a single power law, (2) \feii\ emission, (3) host galaxy.
These three components, which when combined together forms the
pseudo-continuum%
\footnote{
Note that the pseudo-continuum here is not defined in the ordinary way, in
which the host galaxy component is not included.
}%
, are plotted in blue. (4) \hb\ emission line plotted in magenta. (5) Broad
\heii\ $\lambda$4686 emission line plotted in cyan. (6) Narrow emission lines
plotted in orange, including \oiii\ $\lambda\lambda$4959, 5007, \heii\
$\lambda$4686, \hei\ $\lambda$4471, and several coronal lines. The summed
model is plotted in red. The bottom panel shows the residual spectrum. Note
that, although \hei\ $\lambda$5876 is not in the fitting window, its profile
is well recovered after removing the \nai\ $\lambda\lambda$5890, 5896 (Na D)
absorption lines from the host galaxy.

Limited by the signal-to-noise ratio (S/N) for the individual-night spectra,
there are degeneracies between several pairs of spectral components,
including: (a) the AGN power-law continuum and the host galaxy; (b) \feii\
emission and the broad \heii\ line; (c) \feii\ emission and the coronal lines.
Thus, we first fit the high-S/N \textit{mean spectrum} with all parameters set
free. We then fit each individual-night spectrum with the values of some
parameters fixed to those obtained from the fitting of the mean spectrum. The
details of each spectral component, and the fitting parameters, are described
in the following subsections.

\subsubsection{AGN Power-law Continuum}

A single power law is used to describe the featureless AGN continuum. It has
two parameters: the flux density at 5100 \AA\ (\fagn), and the spectral index
($\alpha$, defined as $f_\lambda \propto \lambda^\alpha$). There is some
degeneracy between the power-law continuum and the host galaxy. A larger
$\alpha$ or a higher-flux galaxy (\fgal) both make the total spectrum redder
(note that the unphysical change in the color of the observed spectrum caused
by weather or differential atmospheric refraction is avoided in our
observation, as our flux calibration by the comparison star provides
differential spectrophotometry in each wavelength bin of the spectrum; see
Section \ref{sec-reduction}). 

For our observations, we find that the \textit{relative flux} of the galaxy
component is the main reason for the change in the total-spectrum slope.
Figure \ref{fig-spec382} illustrates this point. The right panel shows the
spectrum of Mrk 382 taken at JD 2456299, just one day after the date of the
spectrum shown in the left panel. Apparently, the flux at 5100 \AA\ is
$\sim$30\% higher than the day before, and the color is redder. However, the
difference between the $V$-band magnitudes of the two nights is only
$\sim$0.01 mag. We have therefore adopted an approach based on the assumption
that the absolute flux of the host must be constant but the relative flux
inside the slit can vary depending on the observing conditions. This is
illustrated in detail in Appendix \ref{sec-host}. The conclusion about the AGN
continuum is in line with several, but not all studies discussing the relation
between $\alpha$ and luminosity in AGNs. Most recently, \citet{zhang13b}
investigated this issue using the reverberation mapping data of the 17 quasars
from \citet{kaspi00}. They found no strong dependence of $\alpha$ on the
variability of the luminosity. Thus, in the fitting of each individual-night
spectrum, we fix the value of $\alpha$ to agree with the best fit to the mean
spectrum, and let \fagn\ and \fgal\ free. Comparing the left and right panels
of Figure \ref{fig-spec382}, only the galaxy component changes. In this way,
the resultant \fagn\ light curve matches the $V$-band light curve well (see
Appendix \ref{sec-host}). Our fitting is consistent with no changes in
spectral index as a function of luminosity.

\subsubsection{\feii\ Emission}

There are several optical \feii\ templates available in the literature. Among
them, two templates, from \citet{boroson92} and \citet{veron04}, are most
widely used, both constructed from the spectrum of the NLS1 galaxy I Zw 1.
\citet{barth13} find that the template from \citet{veron04} yields
inconsistent \hei\ emission lines; significant broad \hei\ $\lambda$4922 and
$\lambda$5016 emission lines are needed while \hei\ $\lambda$4471 is
constrained to have zero flux by the fitting. We compared these two templates
and found that the one from \citet{boroson92} gives better fitting as judged
by the smaller reduced \ks. No broad \hei\ $\lambda$4922 and $\lambda$5016
emission lines are needed, which is consistent with the zero flux of \hei\
$\lambda$4471. Introducing the two broad \hei\ lines will add more parameters
to the fit and will introduce additional degeneracy between \hei\
$\lambda$4922 and \hb. Given this, we chose the \feii\ template from
\citet{boroson92} in this paper.

The \feii\ template is convolved with a Gaussian function to be scaled,
broadened, and shifted \citep[see][for details]{hu08b}. Three parameters, the
flux (\ffe), the full width at half-maximum (\fwhmfe), and shift (\vfe), are
used for \feii\ emission in the fitting. We let all the three parameters to be
free in the fitting of each individual-night spectrum. Note that, \ffe\ is
defined as the flux of the integrated \feii\ emission between 4434 and 4684
\AA\ from the best-fit \feii\ model, following \citet{boroson92}.

\subsubsection{The \hb\ Line}
\label{sec-fithb}

The spectral resolution of our spectra is rather low ($\sim$500 \kms), and in
addition our objects are selected to have narrow \hb\ emission lines. Thus, it
is hard to decompose the \hb\ emission line and remove the narrow component
that comes from the narrow-line region. Considering that the contribution of
the narrow component is weak and supposed to be constant over reverberation
timescales \citep{peterson13}, we treat the entire \hb\ emission line as one
component. The entire \hb\ profile is modeled by a Gauss-Hermite function
\citep{vandermarel93}. All the five parameters of the Gauss-Hermite function
are set free in the fitting of each individual-night spectrum. The flux of the
line (\fhb) is calculated from the best-fit model.

\subsubsection{Narrow Emission Lines}

\begin{figure}
  \centering
  \includegraphics[angle=-90,width=0.45\textwidth]{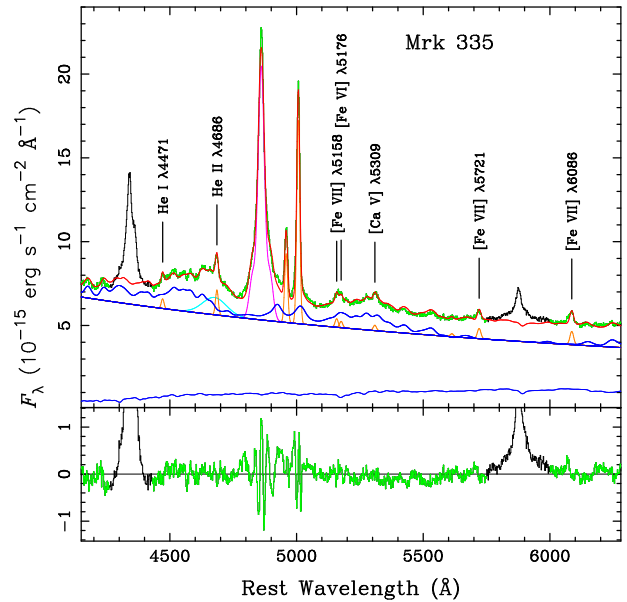}
  \caption{
  An example of individual-night spectrum and model fitting for Mrk 335. The
  spectrum, model, and residuals are plotted in the same manner as that in
  Figure \ref{fig-spec382}. Note the strong narrow \heii\ and coronal lines
  labeled by ion and wavelength. 
  }
  \label{fig-spec335}
\end{figure}

Besides the strong \oiii\ $\lambda\lambda$4959, 5007, there are many other
narrow emission lines in the wavelength range of our fit spectrum
\citep{vandenberk01}. We identify narrow \heii\ $\lambda$4686, \hei\
$\lambda$4471, and several high-ionization forbidden coronal lines. In some
objects, these narrow lines are so strong that their contamination is
non-negligible to the \feii\ measurement. Figure \ref{fig-spec335} shows an
individual-night spectrum of Mrk 335 and our fit. The strong coronal lines
include in this case \fefvii\ $\lambda$5158, \fefvi\ $\lambda$5176, \nfi\
$\lambda$5199, \cafv\ $\lambda$5309, \fefvii\ $\lambda$5721, and \fefvii\
$\lambda$6086. Adding these lines significantly improve the fit.

The intensities of the narrow lines differ from one object to the next. For
each object, we first identify narrow lines in the mean spectrum by testing
the goodness of fit. Only lines that are identified in the mean spectrum are
included in the fitting of individual-night spectra. Each narrow emission line
is modeled by a Gaussian. These lines are coming from the narrow-line region
and hence do not vary on the campaign time scale. Thus, we constrain them to
have the same velocity width and shift with those of the  \oiii\ $\lambda$5007
line. The flux ratio relative to \oiii\ $\lambda$5007 is also kept constant,
as given by the best fit to the mean spectrum (the flux ratio of \oiii\
$\lambda$4959 to $\lambda$5007 is fixed to the theoretical value of 1/3), in
order to avoid the degeneracy between these lines and \feii. Only the flux,
velocity width, and shift of \oiii\ $\lambda$5007 are set free.

\subsubsection{Broad \heii\ Emission Line}

The broad \heii\ $\lambda$4686 emission line is strong in the spectrum of some
objects in our sample. Figure \ref{fig-spec335} shows an example. The line,
plotted in cyan, is much broader than \hb, and strongly blueshifted. Such a
line profile is common for high-ionization lines in NLS1s and consistent with
that of the ultraviolet (UV) \heii\ $\lambda$1640 line in SDSS spectra
\citep{richards11}. It is even more prominent in the root-mean-square (rms)
spectra of our objects, indicating large variability similar to the results in
previous \heii\ reverberation mapping studies
\citep[e.g.,][]{bentz10,barth11,grier12}.

The broad \heii\ line is heavily blended with \feii\ emission. The
high-ionization lines in AGN are often asymmetric \citep{richards11}, but
there is no other high-ionization line in the wavelength range of our spectra
to constrain the profile of \heii\ line. We first fit the line profile in the
mean spectrum using a single Gaussian. We then kept the width and the shift in
the individual-night spectra but leave the line intensity as a free parameter. 

The intensity variation of the \heii\ lines in our sample will be discussed in
a forthcoming publication. Here we only note that the procedure described
above seems to be consistent with the observations, and, in general, the lag
of this line is much shorter than the \hb\ lag.

\subsubsection{Host Galaxy}

Following \citet{barth13}, we use single simple stellar population models from
\citet{bruzual03} as templates for the galaxy component. For most objects in
our sample, the instantaneous-burst model with an age of 11 Gyr and solar
metallicity ($Z = 0.02$) provided sufficiently good fit to the mean spectrum
and consistent flux ratio compared with the measured \textit{Hubble Space
Telescope} (\textit{HST}) images (see Papers I and II). In a few cases (Mrk
335, Mrk 142, and Mrk 42), this template gives a flux that is much larger
than that derived from the \textit{HST} image. The template with 11 Gyr and $Z
= 0.05$ provides better results and was adopted in these cases. 

Like the \feii\ template, the galaxy template is also convolved with a
Gaussian to be scaled, broadened, and shifted. As described previously, and
detailed in Appendix \ref{sec-host}, the relative flux \fgal\ is free to vary
in the fitting of individual-night spectra.

\subsection{Fitting Results}

Table \ref{tab-fitfix} lists the values of the parameters fixed in the fitting
of individual-night spectra for all the objects in our sample. Columns (5) to
(12) list the narrow emission lines included in the fits. As explained, their
velocity widths and shifts are constrained to be the same with those of the
\oiii\ $\lambda$5007 line while the relative intensity ratios are kept as
measured from the mean spectrum. The last column gives the galaxy template
used. In the table, a blank entry means that spectral component is not
included for the object.

\begin{deluxetable*}{lccccccccccccc}
  \tablewidth{0pt}
  \tablecolumns{14}
  \tablecaption{Parameters Fixed in the Fitting of Individual-night Spectrum
  \label{tab-fitfix}}
  \tablehead{
  \colhead{Object} & \colhead{Power Law} & \multicolumn{2}{c}{Broad \heii\
  $\lambda$4686} & \colhead{} & \multicolumn{8}{c}{Narrow Emission Lines} &
  Galaxy model
  \\ \cline{3-4} \cline{6-13}
  \colhead{} & \colhead{$\alpha$} & \colhead{FWHM\tablenotemark{a}} &
  \colhead{Shift} & \colhead{} & \colhead{\hei} & \colhead{\heii} &
  \colhead{\fefvii} & \colhead{\fefvi} & \colhead{\nfi} & \colhead{\cafv} &
  \colhead{\fefvii} & \colhead{\fefvii} & \colhead{}
  \\ 
  \colhead{} & \colhead{($f_\lambda \propto \lambda^\alpha$)} &
  \colhead{(\kms)} & \colhead{(\kms)} & \colhead{} &
  \colhead{$\lambda$4471} & \colhead{$\lambda$4686} & 
  \colhead{$\lambda$5158} & \colhead{$\lambda$5176} & 
  \colhead{$\lambda$5199} & \colhead{$\lambda$5309} & 
  \colhead{$\lambda$5721} & \colhead{$\lambda$6086} & \colhead{}
  \\
  \colhead{(1)} & \colhead{(2)} & \colhead{(3)} & \colhead{(4)} &
  \colhead{} & \colhead{(5)} & \colhead{(6)} & \colhead{(7)} & \colhead{(8)} &
  \colhead{(9)} & \colhead{(10)} & \colhead{(11)} & \colhead{(12)} &
  \colhead{(13)} 
  }
  \startdata
  Mrk 335 & $-$1.45 & 7082 & $-$620 & & 0.042 & 0.115 & 0.043 & 0.032 &
  \nodata & 0.028 & 0.057 & 0.079 & 11Gyr\_z05 \\
  Mrk 1044 & $-$2.01 & 5161 & $-$942 & & 0.216 & \nodata & 0.117 & 0.219 &
  \nodata & \nodata & 0.205 & 0.225 & 11Gyr\_z02 \\
  \iras   & free\tablenotemark{b} & \nodata & \nodata & &
  \nodata & \nodata & \nodata & 0.095 & 0.042 & \nodata & \nodata &
  \nodata & \nodata \\
  Mrk 382 & $-$1.96 & 5797 & $-$1322 & & 0.028 & 0.074 & \nodata & \nodata &
  \nodata & 0.021 & 0.033 & 0.021 & 11Gyr\_z02 \\
  Mrk 142 & $-$2.11 & 4824 & $-$757 & & 0.041 & 0.131 & 0.055 & 0.116 &
  \nodata & \nodata & 0.082 & 0.095 & 11Gyr\_z05 \\
  \mcg & $-$0.84 & 5191 & $-$927 & & 0.063 & 0.096 & \nodata & \nodata &
  \nodata & \nodata & 0.046 & 0.061 & 11Gyr\_z02 \\
  \irasf\tablenotemark{c} & $-$2.28 & 5506 & $-$334 & & \nodata & 0.034 &
  0.013 & \nodata & 0.018 & \nodata & 0.008 & 0.010 & 11Gyr\_z02 \\
  Mrk 42 & $-$0.70 & 1472 & $-$55 & & 0.069 & 0.093 & \nodata & \nodata &
  \nodata & \nodata & \nodata & \nodata & 11Gyr\_z05 \\
  Mrk 486 & $-$0.76 & 4877 & $-$724 & & \nodata & 0.171 & 0.027 & \nodata &
  \nodata & 0.067 & 0.050 & 0.140 & 11Gyr\_z02 \\
  Mrk 493 & $-$0.91 & 4054 & $-$1580 & & 0.132 & 0.261 & \nodata & \nodata &
  \nodata & \nodata & \nodata & \nodata & 11Gyr\_z02
  \enddata
  \tablecomments{Fixed parameters in the fitting of individual-night spectra.
  Columns (2) to (4) list the absolute values of the parameters. Columns (5)
  to (12) list the relative intensity ratios with respect to \oiii\
  $\lambda$5007 for the narrow emission lines. Column (13) lists the galaxy
  template from \citet{bruzual03}. \nodata\ means that the component is not
  included in the fitting for the specific object.}
  \tablenotetext{a}{The FWHMs listed are after instrumental broadening
  correction.}
  \tablenotetext{b}{The spectral index of the power-law continuum is free to
  vary in the fitting of the \iras\ spectra, see text for details.}
  \tablenotetext{c}{Intrinsic host galaxy extinction is assumed in the fitting
  of \irasf. See text for details.}
\end{deluxetable*}

Figure \ref{fig-spec} shows examples of fittings to individual-night spectra
for the eight objects not presented so far. The notations and colors are as
the same as those in Figure \ref{fig-spec382} for Mrk 382. Notes on two
objects with unusual treatments are as follows.

\begin{figure*}
  \centering
  \includegraphics[angle=-90,width=0.45\textwidth]{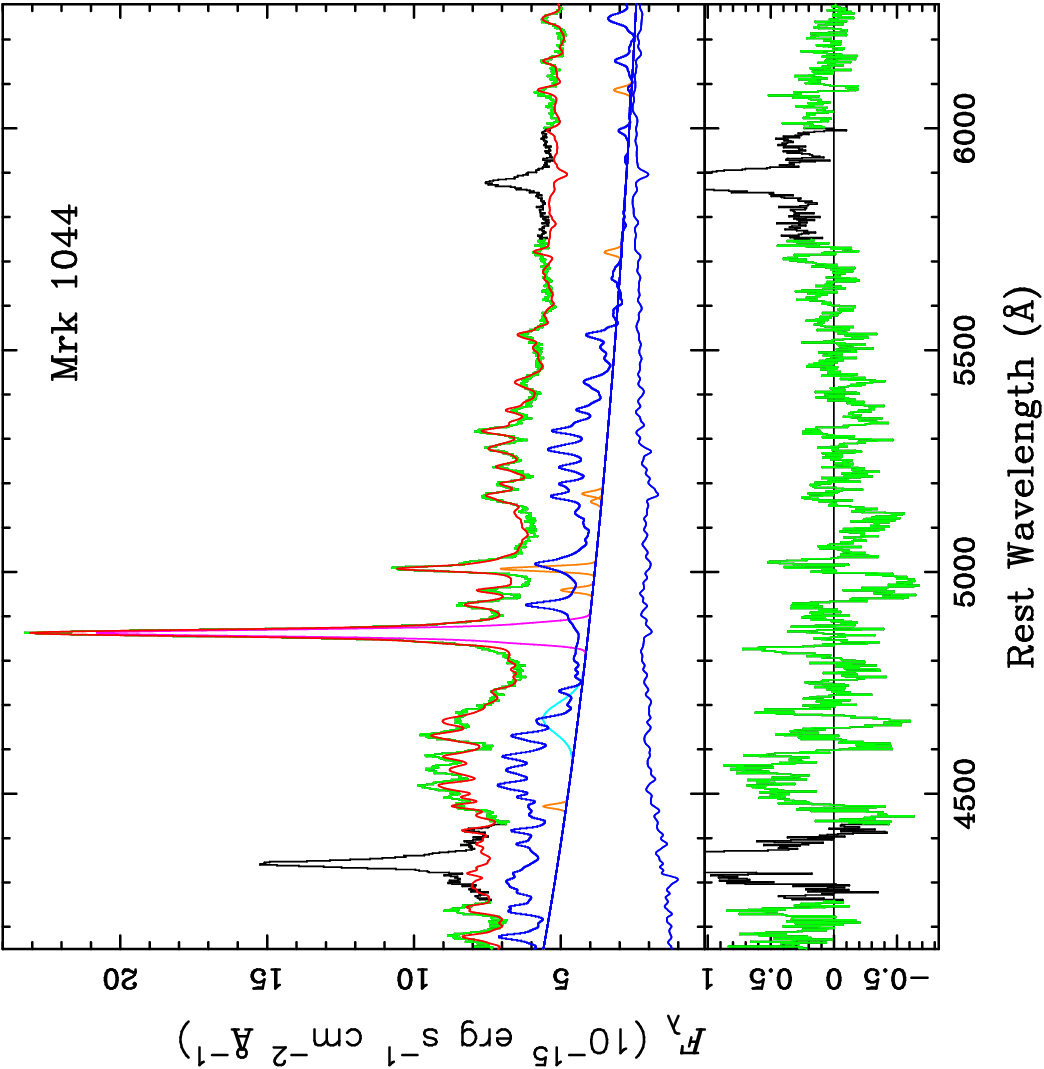}
  \hspace{0.05\textwidth}
  \includegraphics[angle=-90,width=0.45\textwidth]{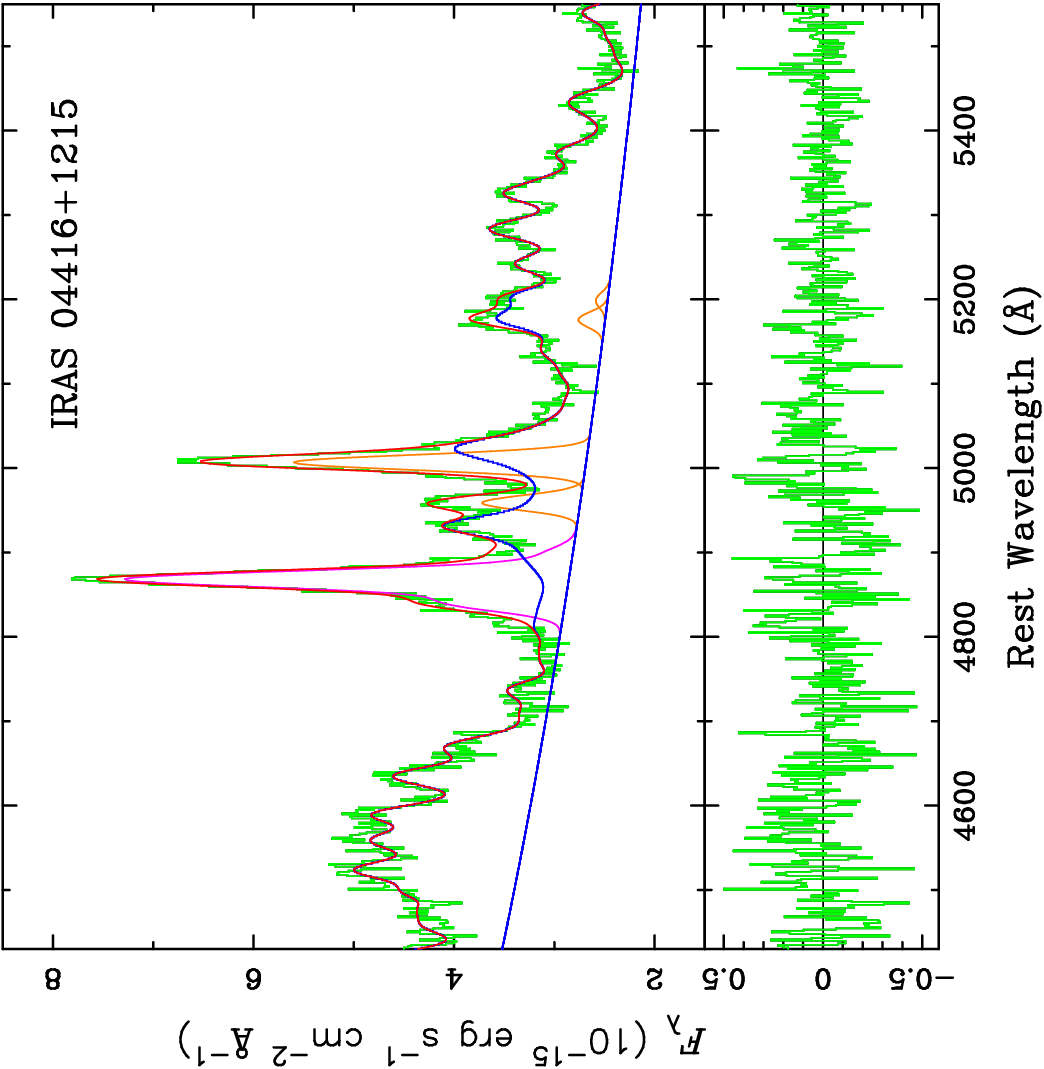}\\
  \vspace{0.025\textwidth}
  \includegraphics[angle=-90,width=0.45\textwidth]{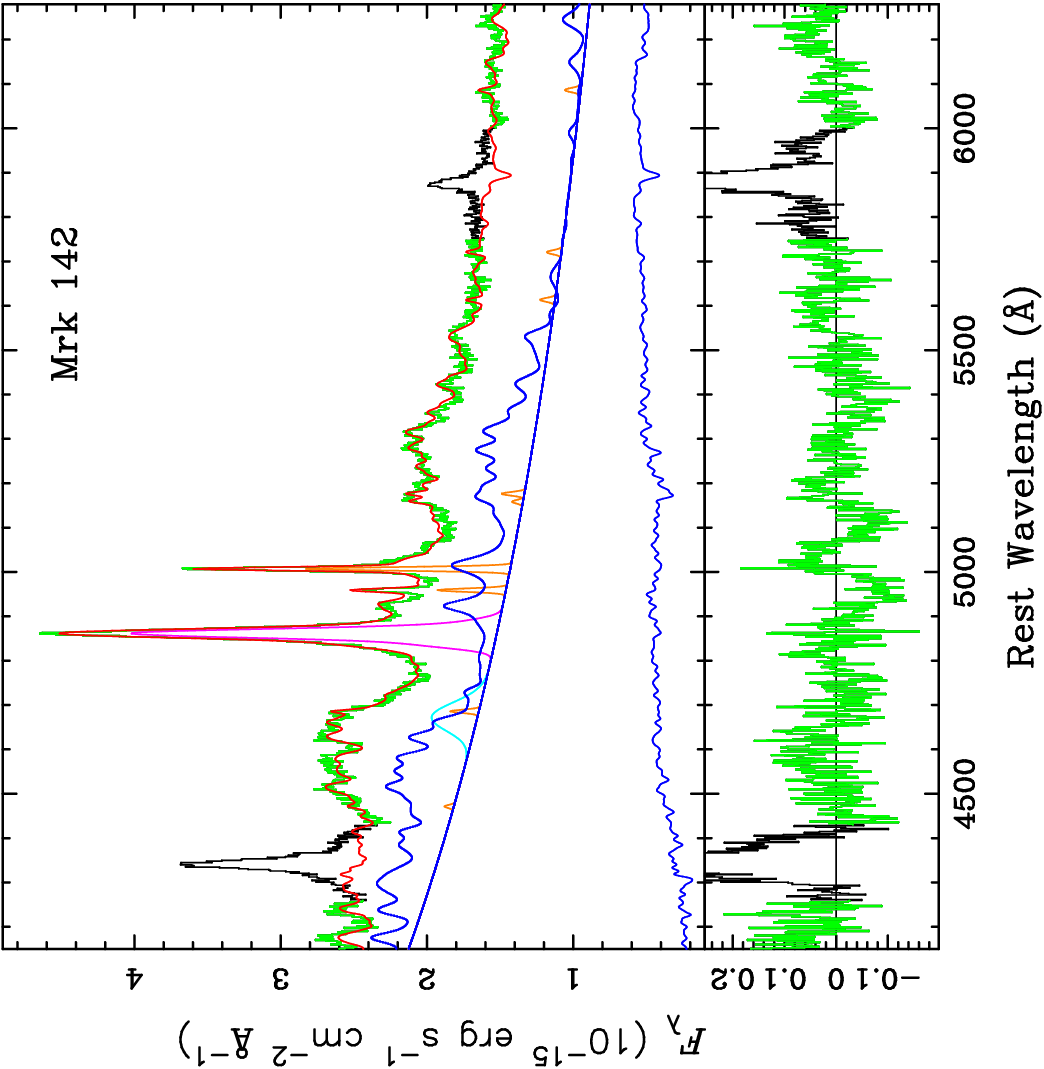}
  \hspace{0.05\textwidth}
  \includegraphics[angle=-90,width=0.45\textwidth]{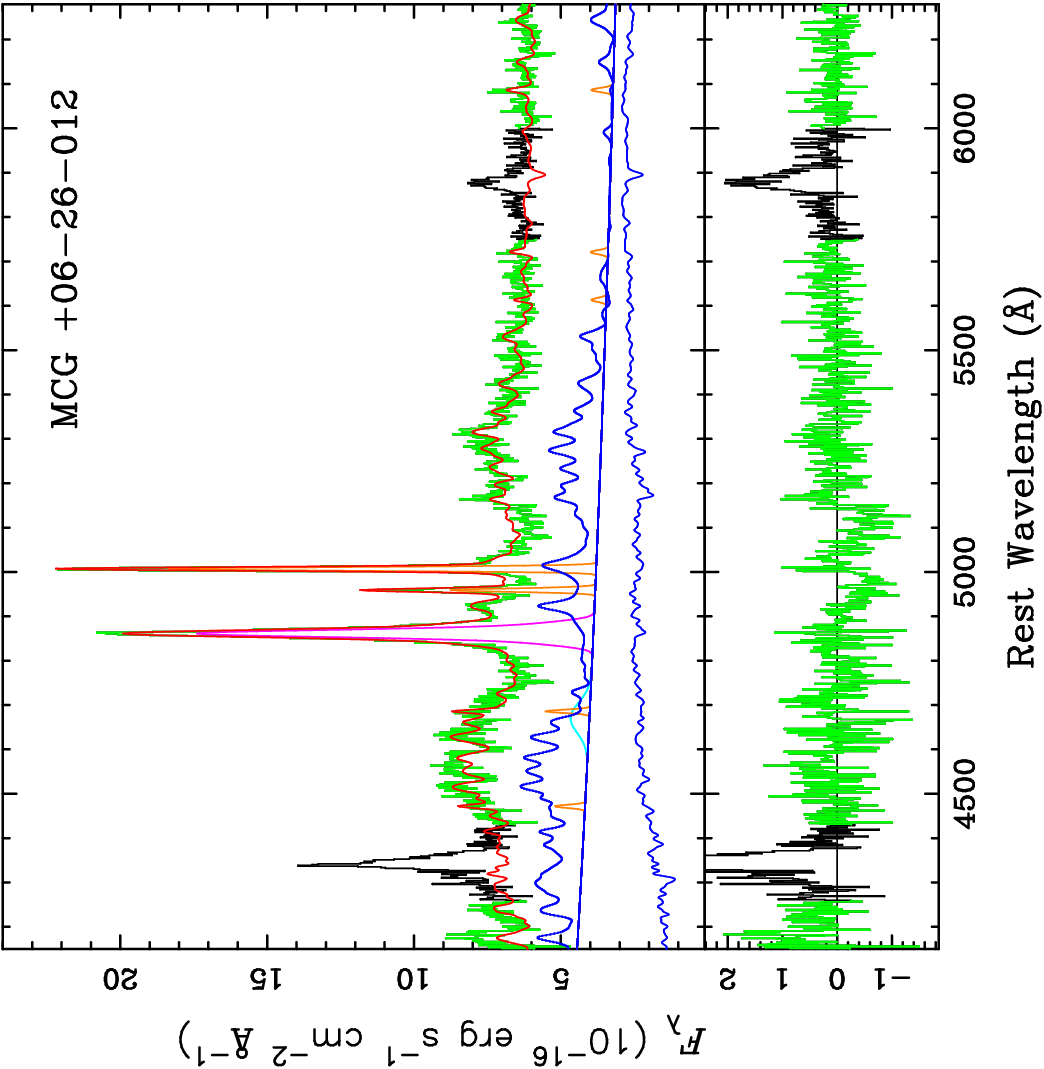}
  \caption{
  Examples of individual-night spectra and model fittings for the other eight
  objects. For each object, the spectrum, model, and residuals are plotted in
  the same manner as those in Figure \ref{fig-spec382}.
  }
  \label{fig-spec}
\end{figure*}

\begin{figure*}
  \figurenum{\ref{fig-spec}}
  \centering
  \includegraphics[angle=-90,width=0.45\textwidth]{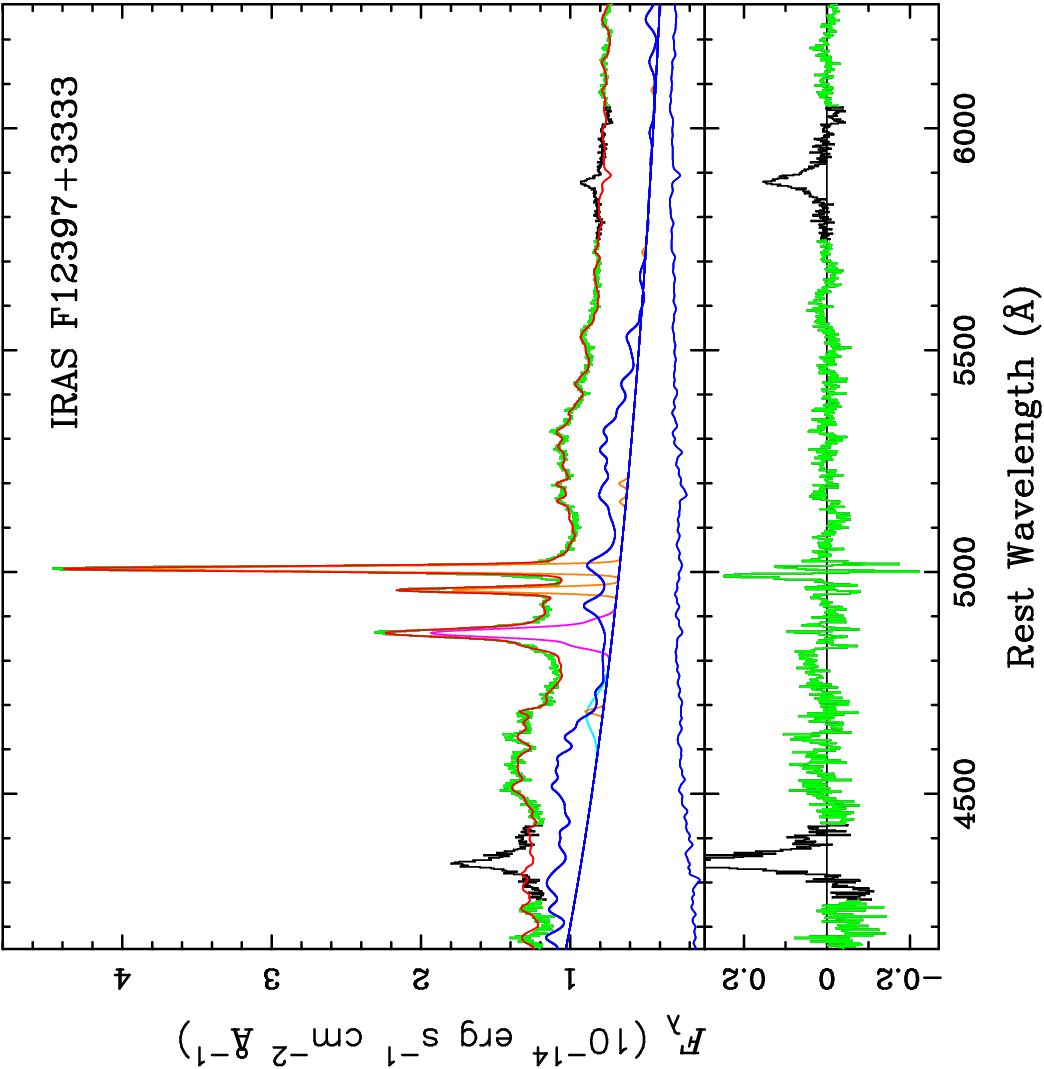}
  \hspace{0.05\textwidth}
  \includegraphics[angle=-90,width=0.45\textwidth]{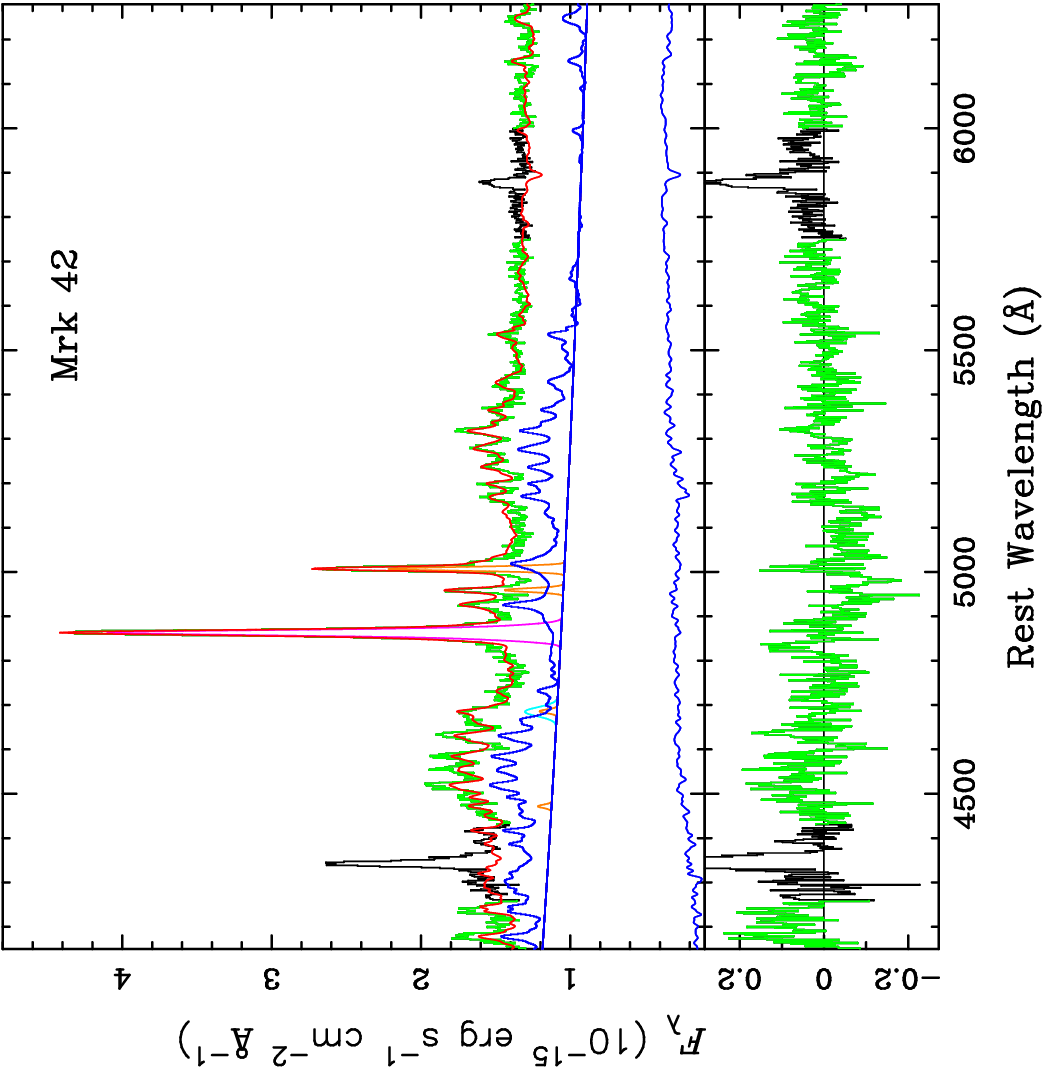}\\
  \vspace{0.025\textwidth}
  \includegraphics[angle=-90,width=0.45\textwidth]{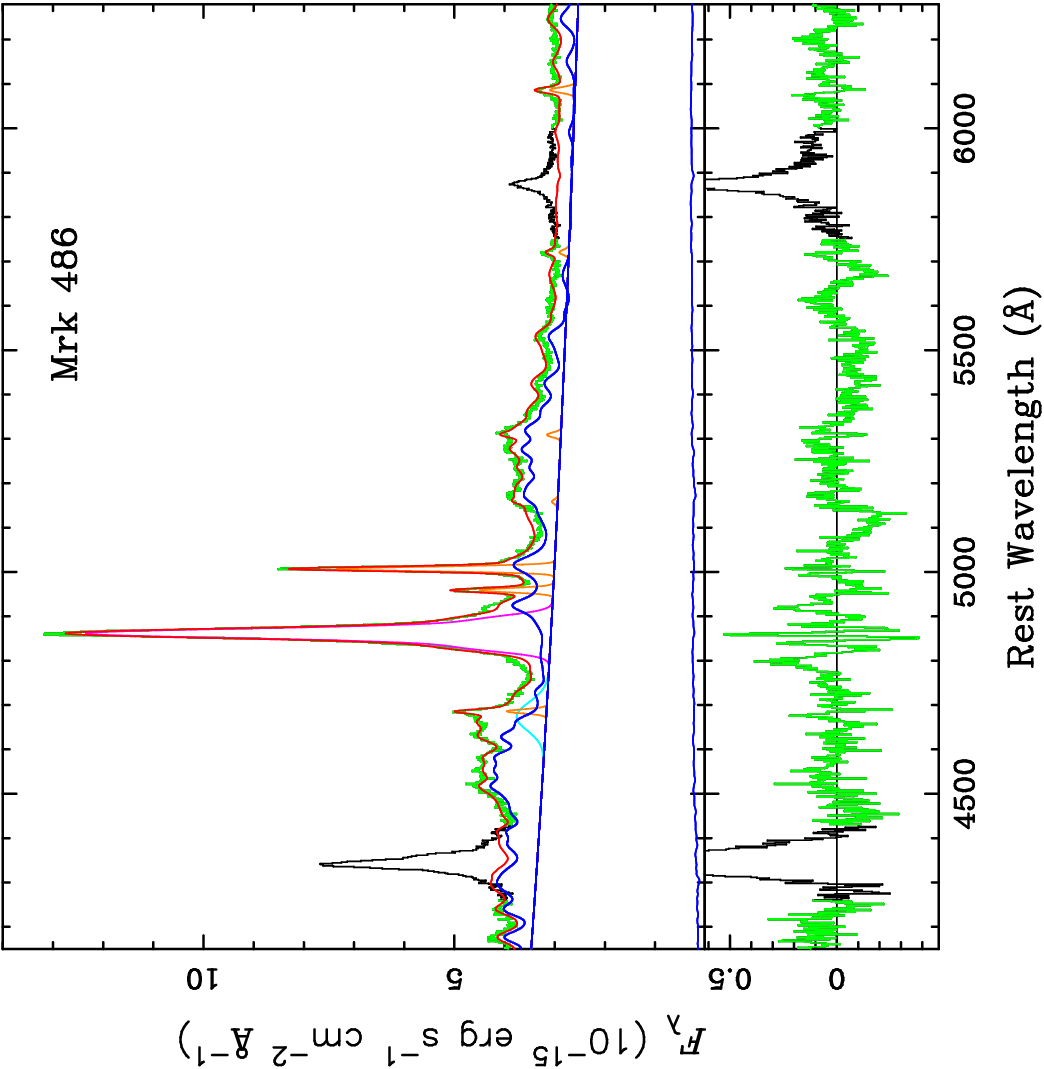}
  \hspace{0.05\textwidth}
  \includegraphics[angle=-90,width=0.45\textwidth]{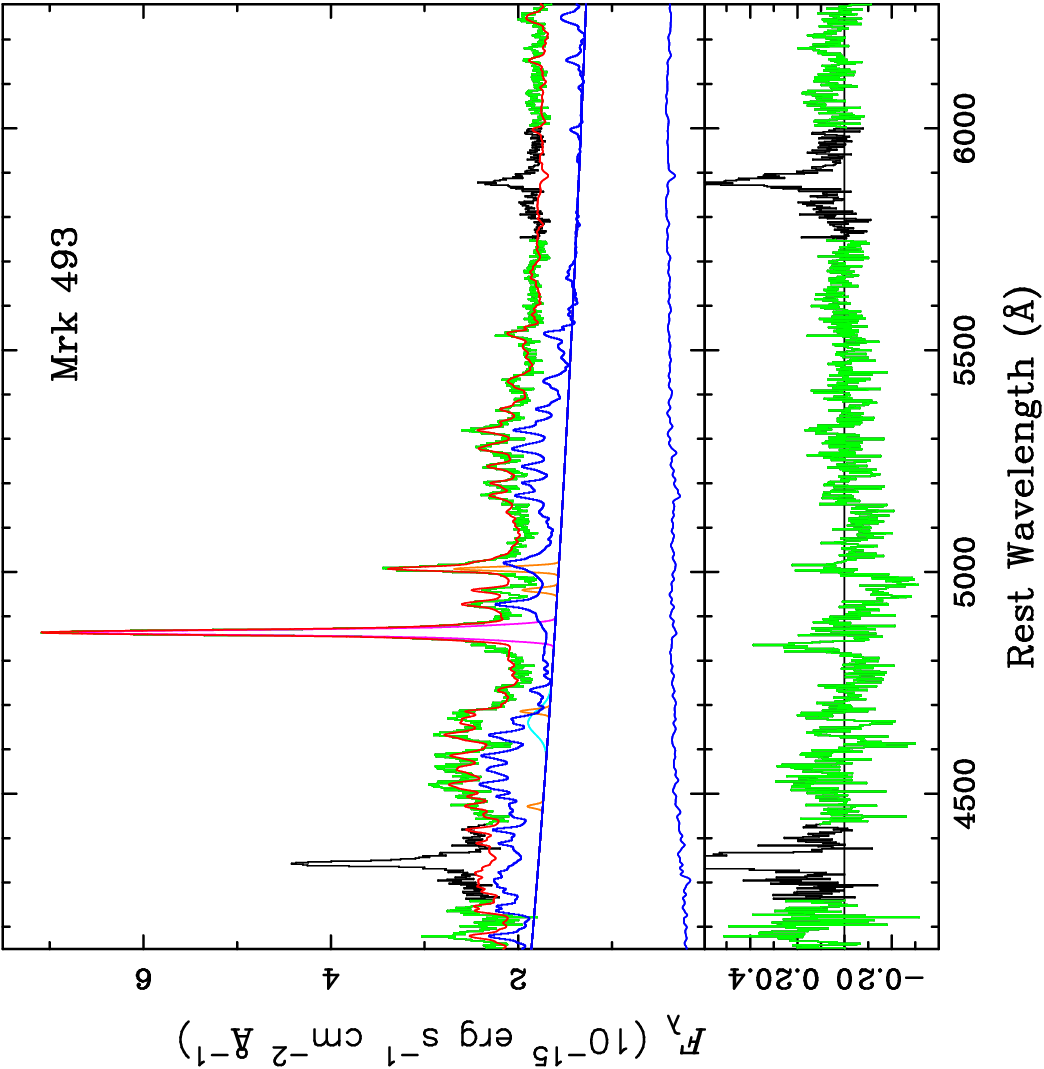}
  \caption{Continued.
  }
\end{figure*}

\textbf{\iras.} This object has the highest redshift (0.089) in our sample.
From the images taken for slit centering, the point spread function of the
object is as broad as that of the comparison star, meaning that there is a
negligible galaxy contribution. For this object, the comparison star is
fainter than the AGN, especially in the red part. We found that the shape of
the spectrum of this object is not as well calibrated by the comparison star
as in the other objects. Thus, we fit in a relatively narrow wavelength window
(4430--5550 \AA), but covering the two bumps of \feii\ emission on either side
of \hb. We also let the spectral index $\alpha$ of the AGN power-law continuum
free to vary in individual-night spectra to compensate for the difference in
spectral shape caused by the calibration. As shown in Figure \ref{fig-spec},
the spectrum can be fitted well by including only a power law, \feii\
emission, \hb, \oiii, and a few narrow lines.

\textbf{\irasf.} As discussed in Paper I, the spectrum of this AGN is probably
affected by host galaxy extinction. The observed spectra (see Figure 4 of
Paper I for an example) have a very red color, and are badly fitted by our
spectral model. We perform an extinction correction after de-redshifting,
assuming the Galactic extinction law and $A_V$ = 1.44 mag estimated from the
Balmer decrement (see details in Paper I). This dereddened spectrum is fit
well with a typical spectral index as shown in Figure \ref{fig-spec}.

\subsection{Line-profile Measurements}

For each individual-night spectrum, the fluxes, FWHMs, and velocity shifts
(with respect to \oiii) of \feii\ and \hb\ are calculated from the best-fit
model. The means of these properties are considered to be the measurements of
the line profile, except for the width of \hb\ (\fwhmhb; see below). The
standard deviations on those means are used as the uncertainties.

\fwhmhb\ is underestimated in the fitting, as we do not include the narrow
component of \hb\ in our model (see Section \ref{sec-fithb}). Thus, we use a
different method to estimate this width. We add a Gaussian to our model to
represent the narrow \hb\ component. The velocity width and shift are
constrained to be the same as those of \oiii. We then fit the mean spectra
twice, once assuming the flux of the narrow \hb\ line is 10\% of the flux of
\oiii\ $\lambda$5007 and once assuming a flux ratio of 0.2. The width of the
broad \hb\ obtained with flux ratio set to 0.1 is adopted as the \fwhmhb. The
uncertainty is obtained from the fit with a flux ratio set to 0.2 and the
original fit that assumed no narrow \hb. This method essentially resembles the
one used in Papers I and II, while the spectral model here is more
sophisticated.

Table \ref{tab-profile} lists the measurements of the 10 objects. The
instrumental broadening (FWHM $\sim$ 500 \kms, Section \ref{sec-fithb}) has
been taken into account in the listed FWHMs. Note that the \hb\ of \iras\ has
large velocity shift with respect to \oiii\ $\lambda$5007, and its profile
shows no feature of a narrow component with the same shift of \oiii. Adding a
narrow component makes the width of the broad \hb\ even narrower. Thus, the
\fwhmhb\ of \iras, as other properties, was obtained from the measurements of
individual-night spectra. 

\begin{deluxetable*}{lr@{~$\pm$~}lr@{~$\pm$~}lr@{~$\pm$~}lcr@{~$\pm$~}lr@{~$\pm$~}lr@{~$\pm$~}l}
  \tablewidth{0pt}
  \tablecolumns{14}
  \tablecaption{Line-profile Measurements
  \label{tab-profile}}
  \tablehead{
  \colhead{Object} & \multicolumn{6}{c}{\feii} & \colhead{} &
  \multicolumn{6}{c}{\hb}
  \\ \cline{2-7} \cline{9-14}
  \colhead{} & \multicolumn{2}{c}{Flux} & \multicolumn{2}{c}{FWHM} &
  \multicolumn{2}{c}{Shift} & \colhead{} & \multicolumn{2}{c}{Flux} &
  \multicolumn{2}{c}{FWHM\tablenotemark{a}} & \multicolumn{2}{c}{Shift}
  \\
  \colhead{} & \multicolumn{2}{c}{($\rm 10^{-15}~erg~s^{-1}~cm^{-2}$)} &
  \multicolumn{2}{c}{(\kms)} & \multicolumn{2}{c}{(\kms)} & \colhead{} &
  \multicolumn{2}{c}{($\rm 10^{-15}~erg~s^{-1}~cm^{-2}$)} &
  \multicolumn{2}{c}{(\kms)} & \multicolumn{2}{c}{(\kms)} 
  }
  \startdata
  Mrk 335      
  &  253 &   9 & 1947 & 143 &   49 &  37 &
  &  661 &  22 & 2096 & 170 &   $-$1 &  16
  \\
  Mrk 1044     
  &  369 &  11 &  866 &  25 &   76 &  10 &
  &  380 &  16 & 1178 &  22 &   12 &  12
  \\
  \iras   
  &  292 &   9 & 1313 &  50 &  496 &  21 &
  &  149 &   4 & 1522 &  44\tablenotemark{b} &  241 &  29
  \\
  Mrk 382      
  &   27 &   4 & 1326 & 234 &   13 & 115 &
  &   39 &   2 & 1462 & 296 &  $-$44 &  35
  \\
  Mrk 142      
  &   87 &   5 & 1512 &  69 &  $-$25 &  37 &
  &   78 &   5 & 1588 &  58 & $-$101 &  28
  \\
  \mcg      
  &   41 &   4 & 1155 &  70 &  $-$21 &  41 &
  &   41 &   4 & 1334 &  80 &  $-$29 &  23
  \\
  \irasf  
  &  550 &  28 & 1748 &  78 &   36 &  40 &
  &  405 &  18 & 1802 & 560 &  $-$50 &  26
  \\
  Mrk 42       
  &   67 &   3 &  787 &  16 &  105 &  19 &
  &   57 &   2 &  802 &  18 &   87 &  19
  \\
  Mrk 486      
  &  186 &   5 & 1790 &  88 &   95 &  33 &
  &  346 &  12 & 1942 &  67 &  $-$46 &   9
  \\
  Mrk 493      
  &  102 &   3 &  780 &   9 &  171 &   9 &
  &   92 &   3 &  778 &  12 &  126 &  13
  \enddata
  \tablecomments{Fluxes, FWHMs, and velocity shifts of \feii\ and \hb. Except
  for the \hb\ FWHMs, the listed values are the means and standard deviations
  obtained from the measurements of individual-night spectra. All listed FWHMs
  include a correction due to instrumental broadening. The velocity shifts are
  with respect to \oiii.}
  \tablenotetext{a}{\hb\ FWHMs are estimated from the mean spectrum of each
  object taking into account the narrow \hb\ component, as explained in the
  text.}
  \tablenotetext{b}{\hb\ FWHM for \iras\ is obtained from individual-night
  spectra. See text for details.}
\end{deluxetable*}

\subsection{Light-curve Measurements}
\label{sec-lc}

Our fitting successfully reduces the scatter in the light curves due to the
influence of the host galaxy contamination; thus, \fagn\ represents the real
AGN continuum much better than the simply integrated 5100 \AA\ flux (\fsumc),
as shown in Appendix \ref{sec-host}. The light curves of the emission lines
are also generated directly from the best-fit values of the corresponding
parameters obtained from the fits of individual-night spectra. The errors of
the fluxes given by the fitting are not large enough to account for the
scatter in the fluxes of successive nights, an additional systematic error is
estimated for each light curve (as in Paper I). This systematic error is added
in quadrature to the fitting error for the calculations of variability
amplitudes and time lags below. Note that our treatment is different from that
of \citet{barth13}, who measured the light curve of \hb\ by integrating the
continuum-subtracted spectra.

\section{Light-curve Analysis and Results}
\label{sec-result}

The left columns of Figure \ref{fig-lc} and Figure \ref{fig-lc42} show the
light curves of \fagn, \fhb, and \ffe\ for nine objects with reliable lag
measurements, and the remaining one (Mrk 42), respectively. In this section,
we will calculate the variability amplitudes and the reverberation lags for
the \hb\ and \feii\ emission lines. We then compare the \hb\ time lags  with
those presented in Papers I and II.

\begin{figure*}
  \centering
  \includegraphics[width=0.45\textwidth]{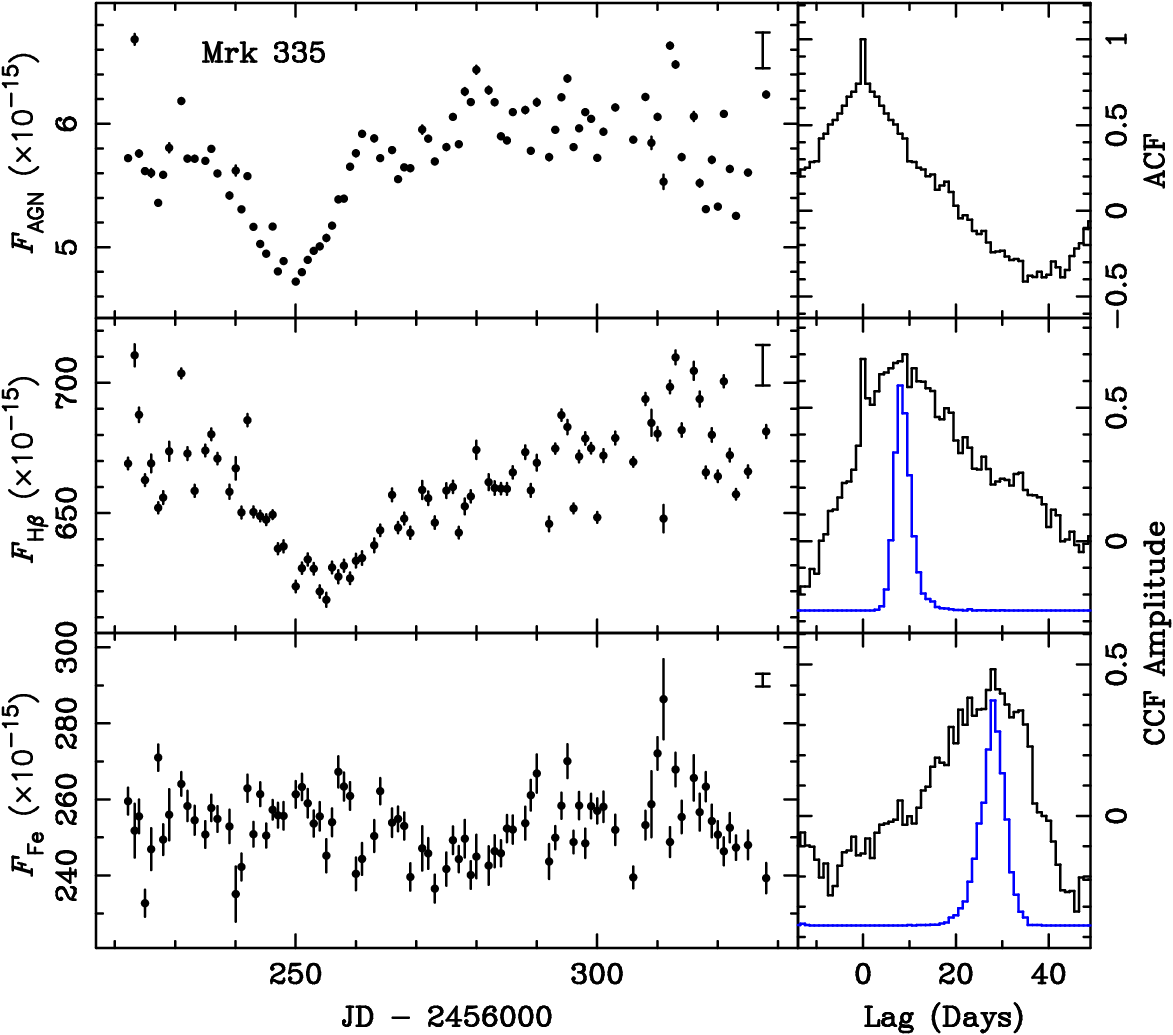}
  \hspace{0.05\textwidth}
  \includegraphics[width=0.45\textwidth]{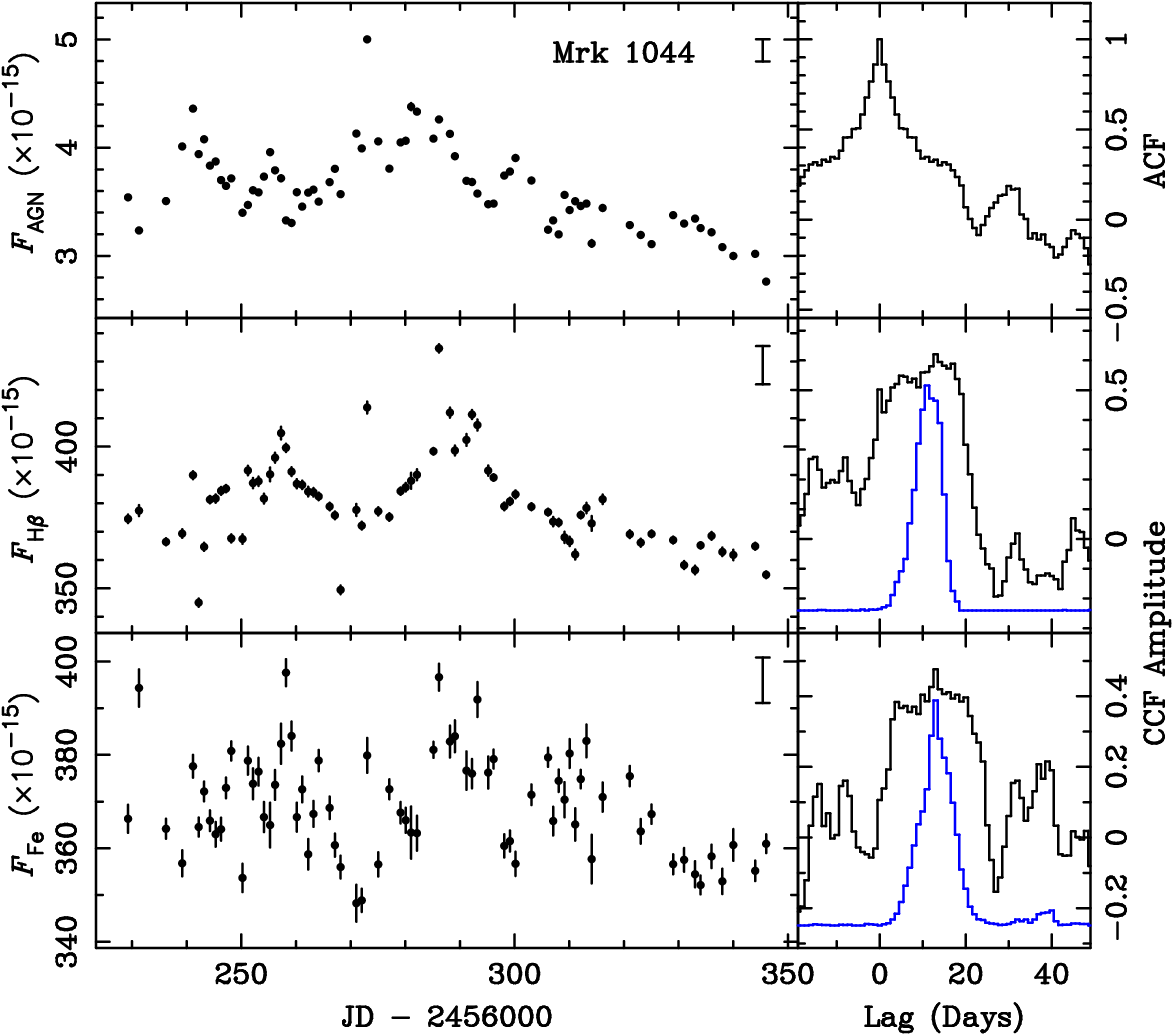}\\
  \vspace{0.025\textwidth}
  \includegraphics[width=0.45\textwidth]{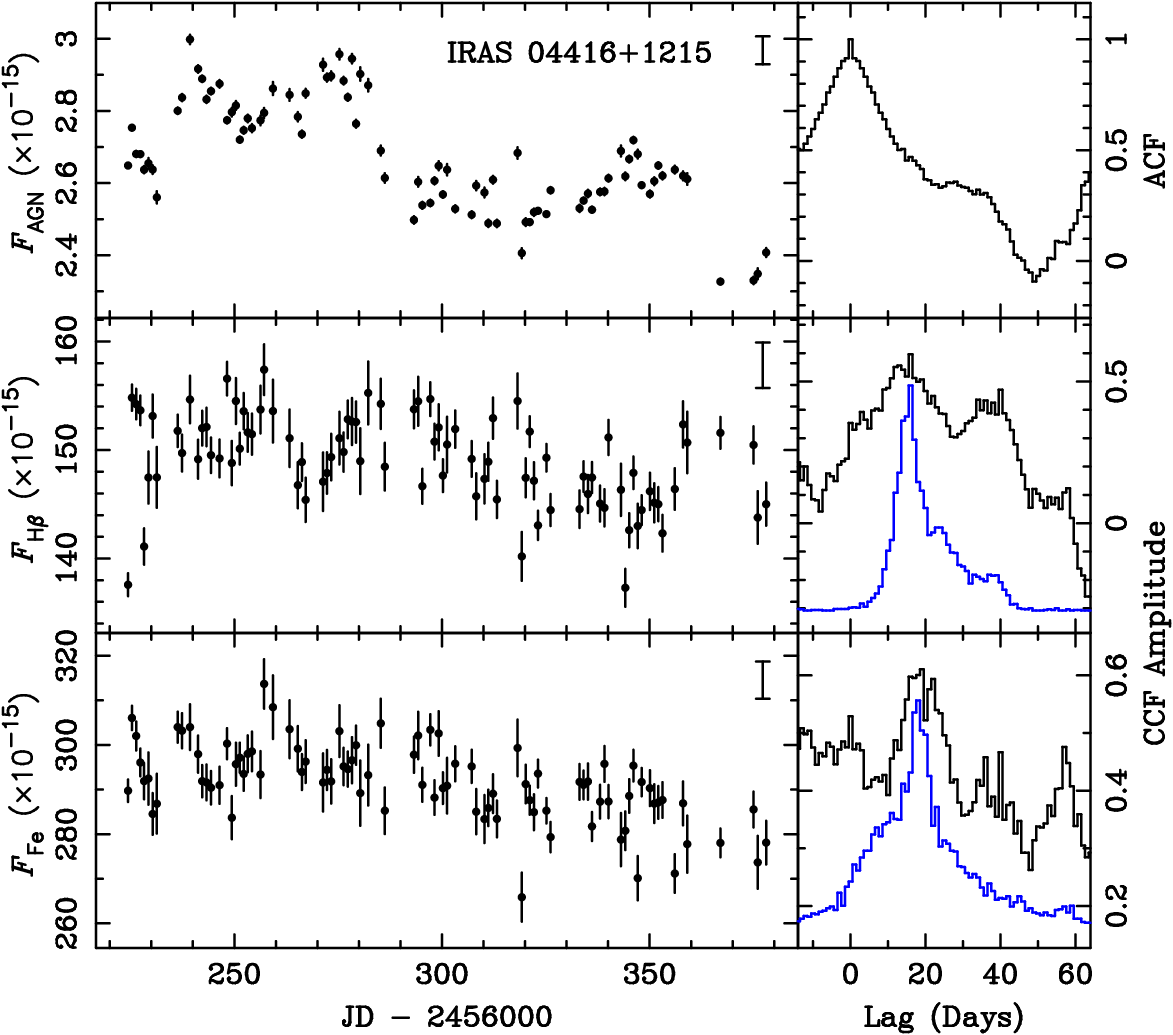}
  \hspace{0.05\textwidth}
  \includegraphics[width=0.45\textwidth]{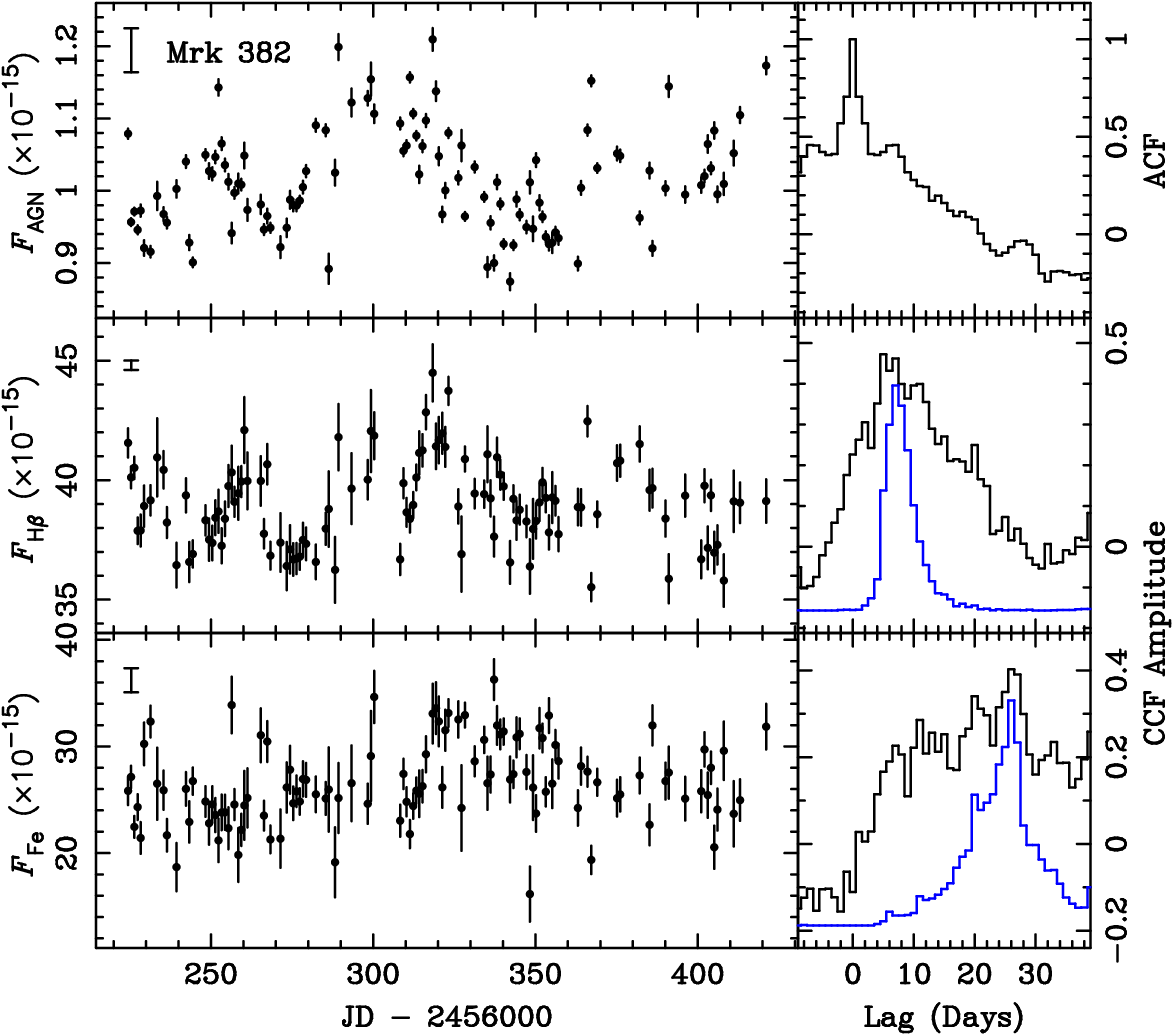}
  \caption{
  Light curves and cross-correlation functions for the nine objects with
  reliable time lag measurements. For each object, the left column show the
  light curves of the AGN (top-left), \hb\ (middle-left), and \feii\
  (bottom-left). The error bar of each flux point is only the fitting error.
  For each light curve, an additional systematic error is estimated from the
  scatter in the fluxes of successive nights, and plotted as the error bar
  with terminals in the corner of the panel. The right column show the
  autocorrelation function of the AGN light curve (top-right),
  cross-correlation functions for \hb\ (middle-right) and \feii\
  (bottom-right) with respect to the AGN continuum. The blue histograms in the
  middle-right and bottom-right panels are the corresponding cross-correlation
  centroid distributions.
  }
  \label{fig-lc}
\end{figure*}

\begin{figure*}
  \figurenum{\ref{fig-lc}}
  \centering
  \includegraphics[width=0.45\textwidth]{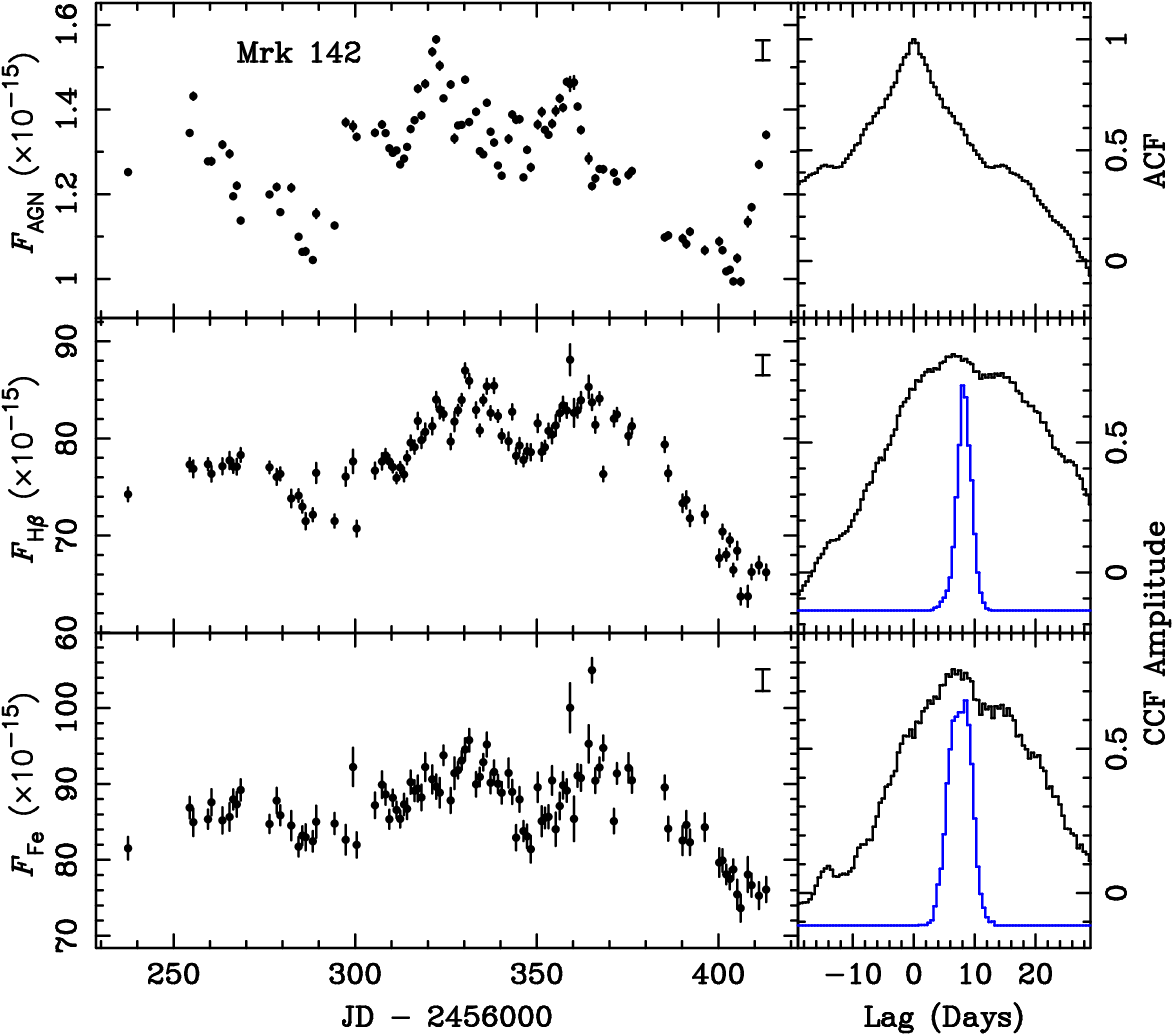}
  \hspace{0.05\textwidth}
  \includegraphics[width=0.45\textwidth]{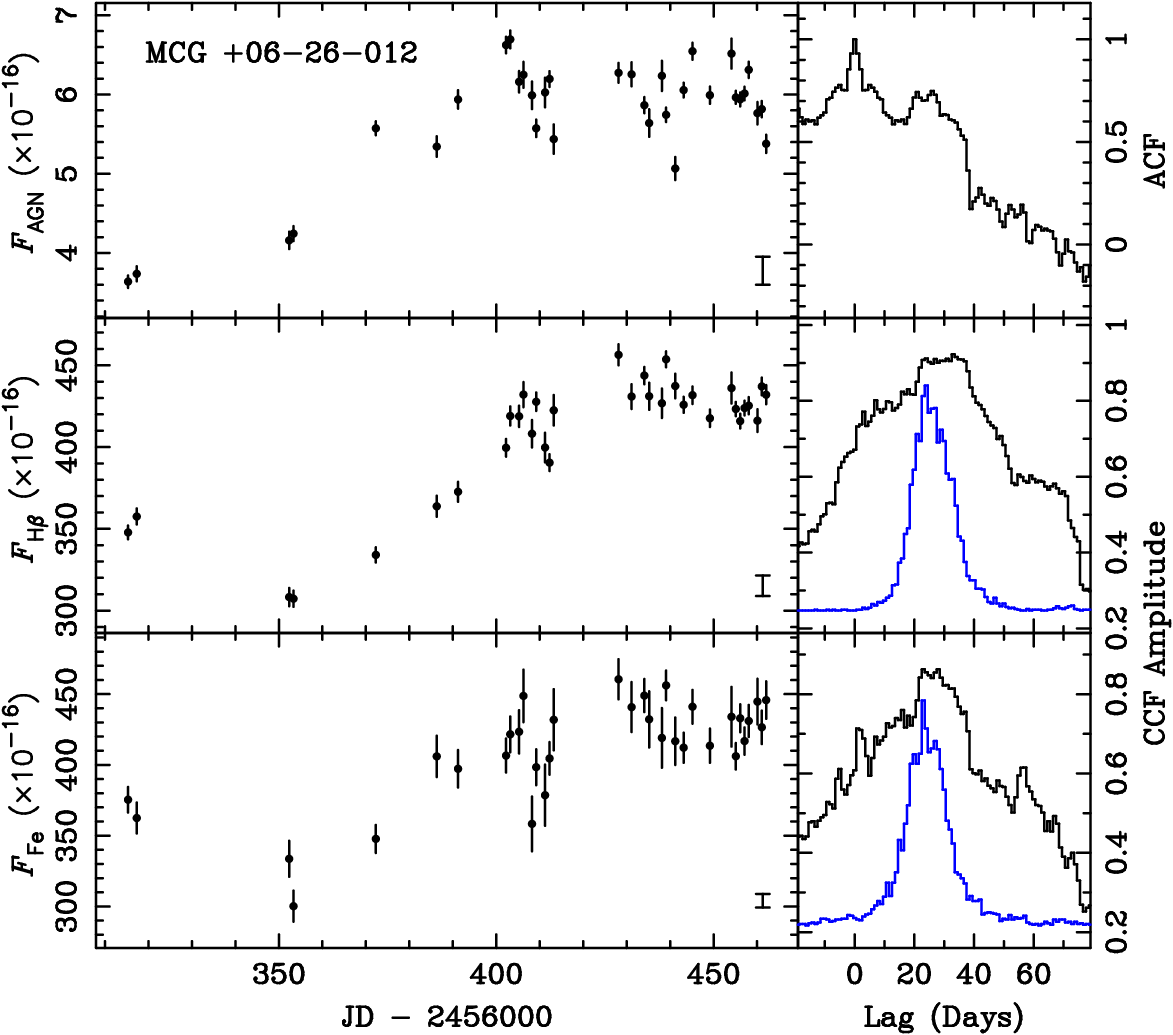}\\
  \vspace{0.025\textwidth}
  \includegraphics[width=0.45\textwidth]{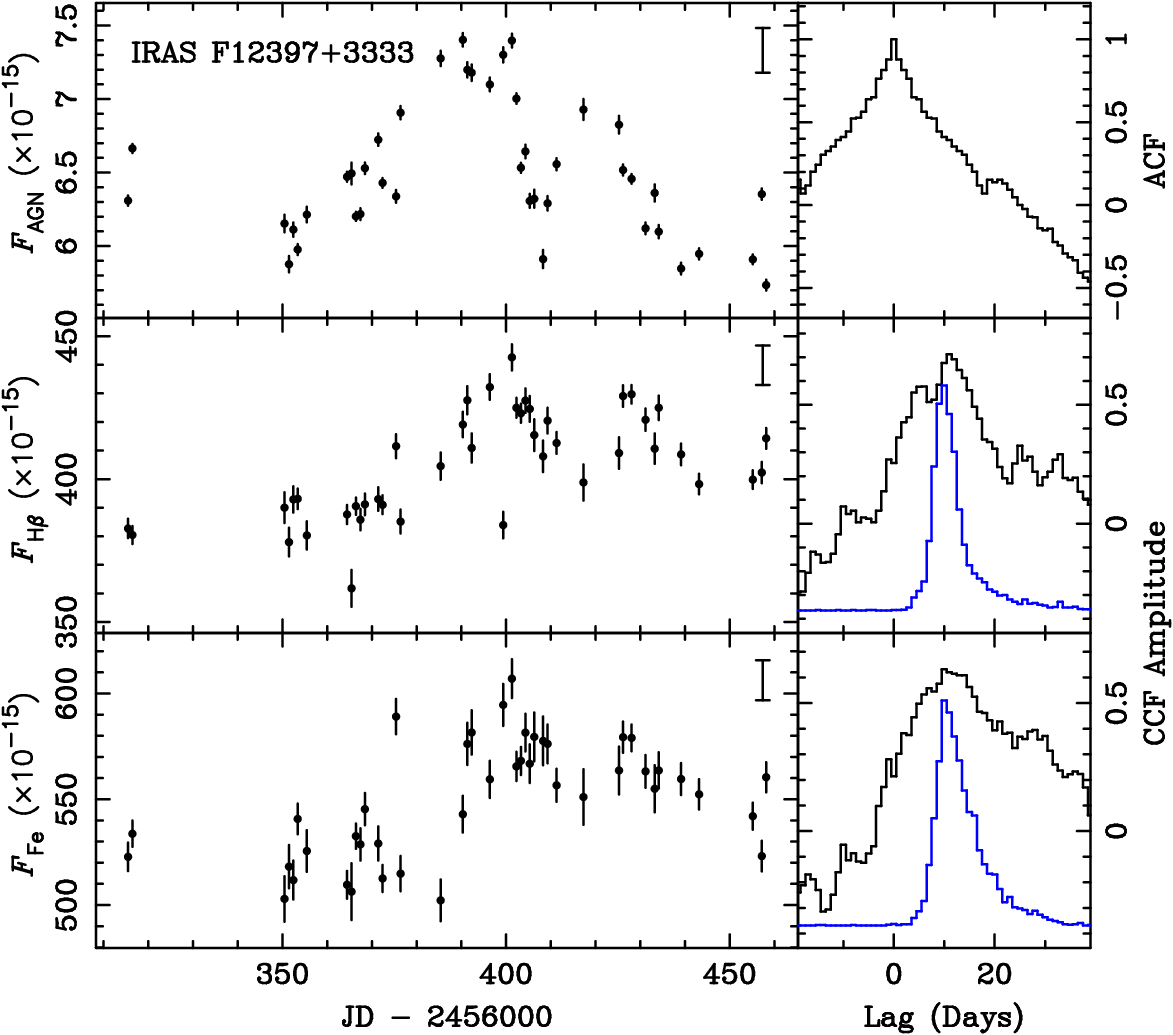}
  \hspace{0.05\textwidth}
  \includegraphics[width=0.45\textwidth]{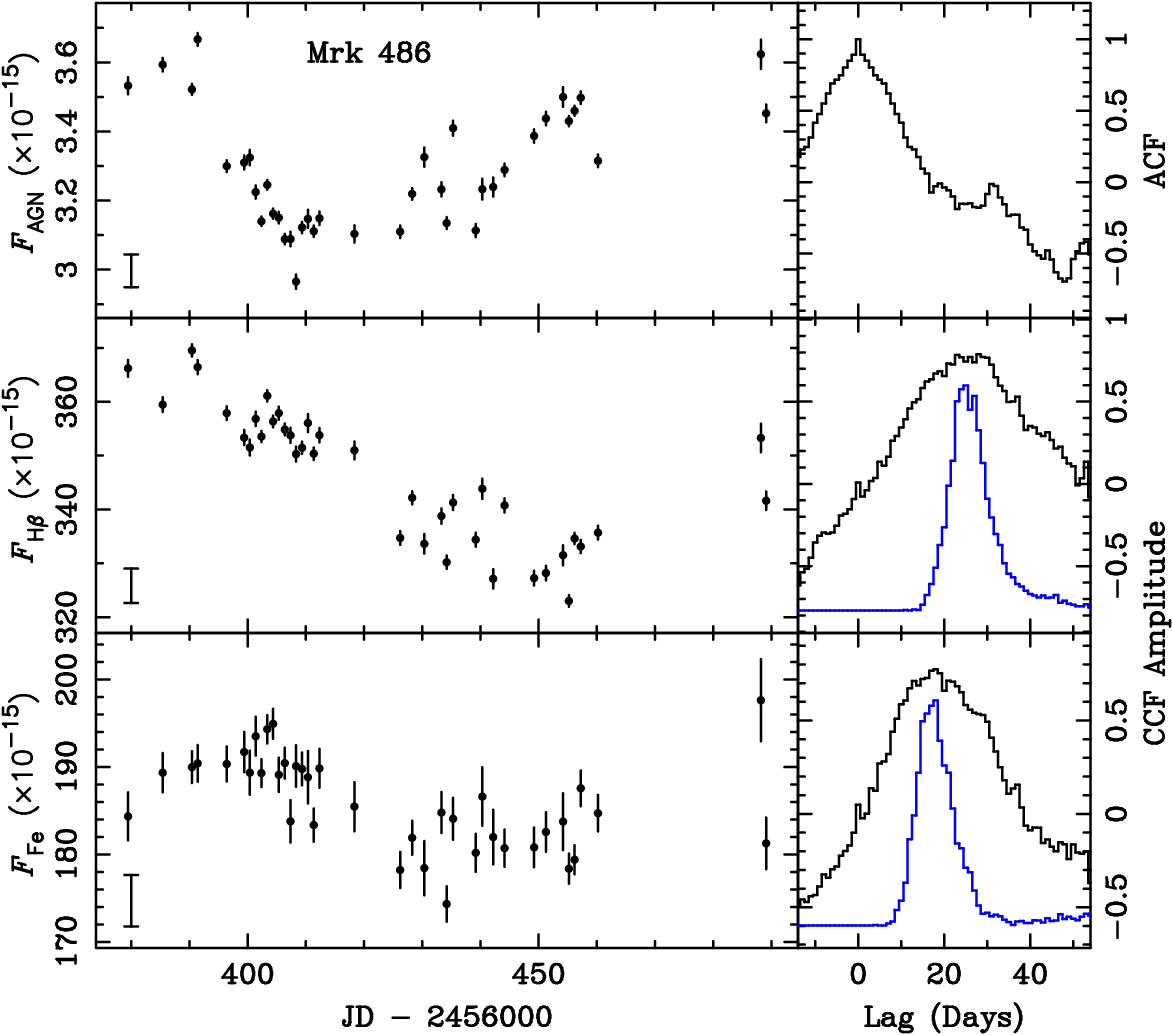}
  \caption{
  Continued.
  }
\end{figure*}

\begin{figure}
  \figurenum{\ref{fig-lc}}
  \centering
  \includegraphics[width=0.45\textwidth]{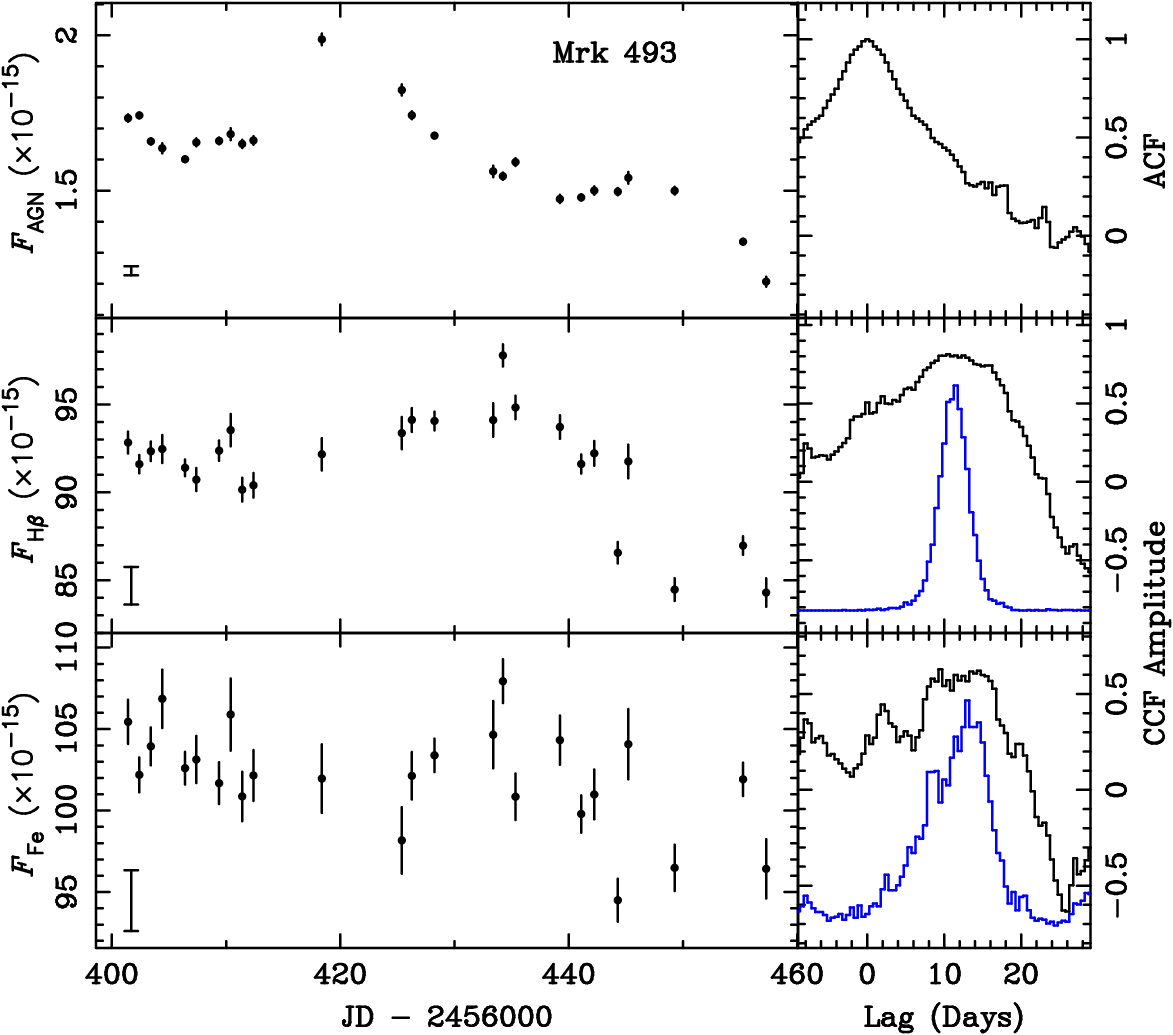}
  \caption{
  Continued.
  }
\end{figure}

\begin{figure}
  \centering
  \includegraphics[width=0.4\textwidth]{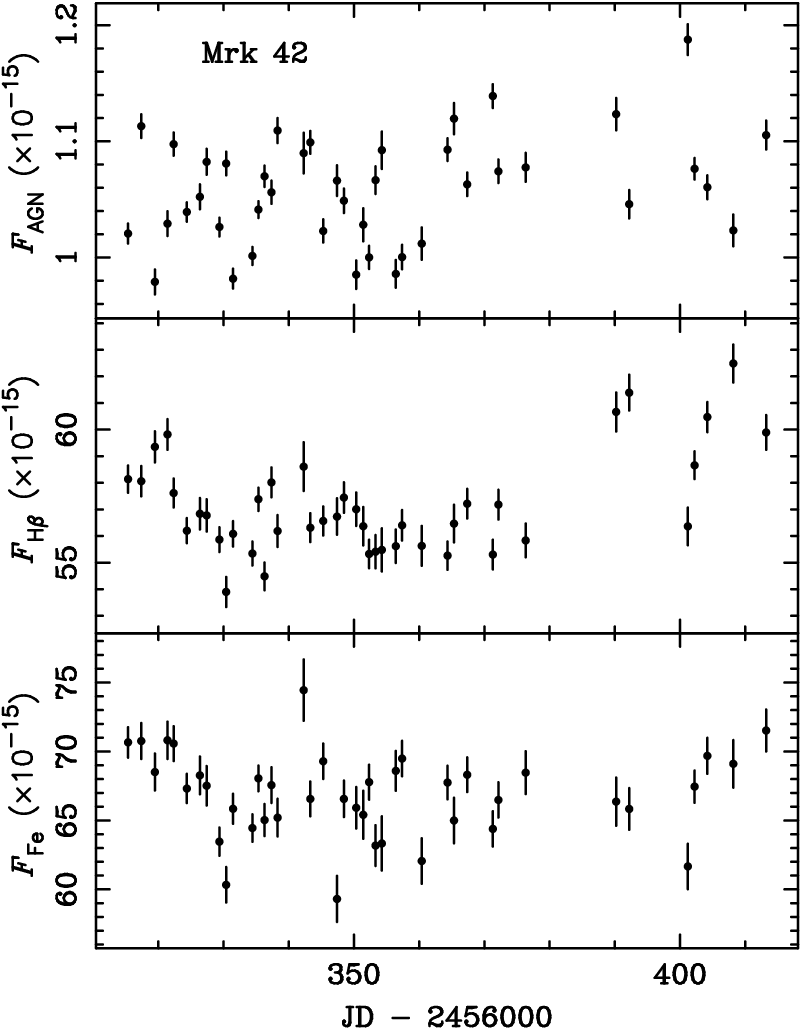}
  \caption{
  Light curves of the AGN (top panel), \hb\ (middle), and \feii\ (bottom) for
  Mrk 42. Note the similarity of the light curves of \hb\ and \feii.
  }
  \label{fig-lc42}
\end{figure}

\subsection{Variability Amplitudes}

We use the quantity \fvar\ defined in \citet{rodriguez97} to represent the
variability amplitude. This quantity is an estimate of the intrinsic
variability over the errors. The uncertainties are calculated following
\citet{edelson02}. The results for \feii\ and \hb\ are listed in Table
\ref{tab-var}. For the variability amplitude ratio of \feii\ to \hb, the range
is from about 0.6 (Mrk 486) to 1.2 (Mrk 42), except for Mrk 382 which has the
largest host galaxy contamination. On average (except for Mrk 382), the \feii\
value is about 10\% smaller, consistent with previous results
\cite[e.g.,][]{vestergaard05,barth13}.

\begin{deluxetable}{lr@{~$\pm$~}lr@{~$\pm$~}l}
  \tablewidth{0pt}
  \tablecolumns{5}
  \tablecaption{Variability Amplitudes
  \label{tab-var}}
  \tablehead{
  \colhead{Object} & \multicolumn{4}{c}{\fvar}
  \\ \cline{2-5}
  \colhead{} & \multicolumn{2}{c}{\feii} & \multicolumn{2}{c}{\hb}
  }
  \startdata
  Mrk 335  &  3.1 &  0.3 &  3.0 &  0.3 \\
  Mrk 1044 &  2.6 &  0.3 &  3.7 &  0.4 \\
  \iras    &  2.1 &  0.3 &  2.0 &  0.3 \\
  Mrk 382  & 11.5 &  1.2 &  4.1 &  0.4 \\
  Mrk 142  &  5.5 &  0.5 &  6.6 &  0.5 \\
  \mcg     &  8.1 &  1.2 &  9.2 &  1.2 \\
  \irasf   &  4.4 &  0.6 &  4.1 &  0.5 \\
  Mrk 42   &  3.6 &  0.7 &  2.9 &  0.4 \\
  Mrk 486  &  2.0 &  0.5 &  3.4 &  0.4 \\
  Mrk 493  &  2.1 &  0.7 &  3.1 &  0.5
  \enddata
  \tablecomments{The listed values are in percentage.}
\end{deluxetable}

\subsection{Reverberation Lags}

The time lags between the AGN continuum variations (\fagn) and the emission
lines (\ffe\ and \fhb) are measured from the cross-correlation functions
(CCFs) for the relevant light curves. We use the interpolation
cross-correlation function \citep[ICCF;][]{gaskell86,gaskell87,white94} method
to calculate the CCF, and adopt the centroid of the CCF, above 80\% of the
peak value (\rmax) as the time lag \citep{koratkar91,peterson04}. The
uncertainty in the time lag measurement is estimated from the
cross-correlation centroid distribution (CCCD) given by random subset
selection/flux randomization Monte Carlo realizations
\citep{maoz89,peterson98}.

All objects in our sample, except for Mrk 42, have reliable time lag
measurements for both \feii\ and \hb. The right columns of Figure \ref{fig-lc}
show the results of the CCF analysis. For each object, the top-right panel
shows the autocorrelation function (ACF) of the \fagn\ light curve, which is
shown in the top-left panel. The two lower panels in the right column show the
CCFs (in black) for \hb\ (middle-right) and \feii\ (bottom-right) with respect
to \fagn\ light curves. The blue histograms are the corresponding CCCDs. Most
CCCDs have a rather symmetric profile, except those for \iras.

Table \ref{tab-lag} lists the time lags of \feii\ (\tfe) and \hb\ (\thb),
their uncertainties, and the corresponding \rmax, for the nine objects with
reliable lag measurements. The listed time lags and uncertainties are in the
rest frame after time-dilation correction. Both time lags for \iras\ have
uncertainties of highly unequal upper and lower limits, as the sequence of
their asymmetric CCCDs. For other objects, the time lags are well determined
by the single-peak CCFs with high \rmax\ and symmetric, narrow CCCDs.

\begin{deluxetable}{lccccc}
  \tablewidth{0pt}
  \tablecolumns{6}
  \tablecaption{Time Lag Measurements
  \label{tab-lag}}
  \tablehead{
  \colhead{Object} & \multicolumn{2}{c}{\feii} & \colhead{} &
  \multicolumn{2}{c}{\hb}
  \\ \cline{2-3} \cline{5-6}
  \colhead{} & \colhead{\rmax} & \colhead{\tfe} & & \colhead{\rmax} &
  \colhead{\thb}
  }
  \startdata
  Mrk 335  & 0.48 & $26.8_{ -2.5}^{+ 2.9}$ & & 0.70 & $ 8.7_{ -1.9}^{+ 1.6}$ \\
  Mrk 1044 & 0.48 & $13.9_{ -4.7}^{+ 3.4}$ & & 0.62 & $10.5_{ -2.7}^{+ 3.3}$ \\
  \iras    & 0.61 & $12.6_{ -6.7}^{+16.7}$ & & 0.60 & $13.3_{ -1.4}^{+13.9}$ \\
  Mrk 382  & 0.40 & $23.8_{ -6.0}^{+ 6.0}$ & & 0.47 & $ 7.5_{ -2.0}^{+ 2.9}$ \\
  Mrk 142  & 0.78 & $ 7.6_{ -2.2}^{+ 1.4}$ & & 0.84 & $ 7.9_{ -1.1}^{+ 1.2}$ \\
  \mcg     & 0.86 & $22.4_{ -6.3}^{+ 9.3}$ & & 0.92 & $24.0_{ -4.8}^{+ 8.4}$ \\
  \irasf   & 0.63 & $10.6_{ -1.9}^{+ 7.0}$ & & 0.71 & $ 9.7_{ -1.8}^{+ 5.5}$ \\
  Mrk 486  & 0.77 & $17.3_{ -3.7}^{+ 5.8}$ & & 0.79 & $23.7_{ -2.7}^{+ 7.5}$ \\
  Mrk 493  & 0.63 & $11.9_{ -6.5}^{+ 3.6}$ & & 0.81 & $11.6_{ -2.6}^{+ 1.2}$
  \enddata
  \tablecomments{Time lags are in the rest frame, and in units of days.}
\end{deluxetable}

The detection rate of \feii\ time lag in our sample is extremely high (9 in
10), compared with previous reverberation-mapping experiments. This is mainly
attributed to the common feature of strong \feii\ emission in the SEAMBH
sample, the high-cadence observation, and sophisticated spectral fitting. The
failure to obtain reliable \tfe\ and \thb\ in Mrk 42 is caused by the \fagn\
light curve, which has large scatter and no clear structure (see Figure
\ref{fig-lc42}). However, the light curves of \ffe\ and \fhb\ show hints for a
similar structure which, unfortunately, is not enough to establish a time lag.

The clear reverberation of the \feii\ emission in response to the continuum
supports the hypothesis that the \feii\ emission originates from photoionized
gas in the BLR. The high detection rate indicates that this is prevalent in
NLS1, which are AGNs with high \er. It is possible that the origin of \feii\
emission in AGNs with low \er\ is different, but there is little evidence to
support this claim. In fact, \citet{barth13} has already detected the
reverberation of \feii\ in two broad-line Seyfert 1 galaxies with \hb\ widths
of FWHM $\gtrsim$ 4000 \kms\ and much lower \er. Comparing the time lags of
\feii\ and \hb, along with their velocity widths, provides some hints about
the location  and geometries of the two emission regions. We will discuss
these ideas in Section \ref{sec-cmpfehb} below. 

\subsection{\hb\ Lag Comparison}

\begin{figure}
  \centering
  \includegraphics[width=0.45\textwidth]{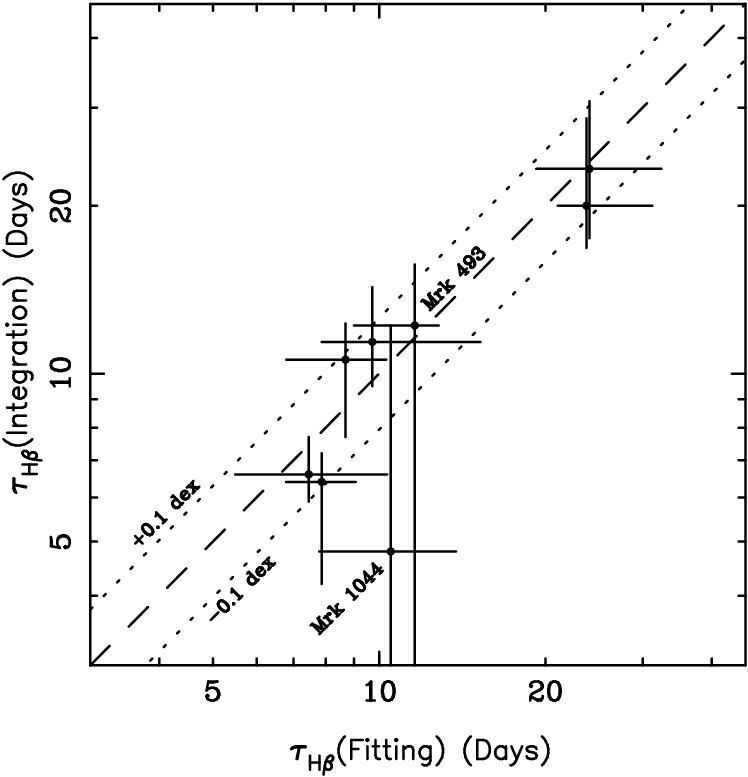}
  \caption{
  A comparison of the \hb\ time lags obtained by simple integration (Papers I
  and II) with those obtained by fitting (this work). The differences are
  typically less than 0.1 dex (dotted lines) except for \iras\ (not plotted
  because its \thb\ cannot be defined by integration) and Mrk 1044 (very large
  uncertainty of the integration method). Mrk 493 also has very large
  uncertainty when measured by the integration method but the measured \thb\
  are very similar.
  }
  \label{fig-hbsumcmp}
\end{figure}

In Paper I and II, the reported time lags of \hb\ are obtained from the light
curves measured by the more traditional integration method, without taking
into account  the contamination by the host galaxy and the narrow emission
lines. Figure \ref{fig-hbsumcmp} shows a comparison between the previously
reported \thb\ based on the integration method and the new results given by
the fitting method reported here. The dashed diagonal line denotes the 1:1
ratio and the region between the two dotted lines gives the 0.1 dex deviation
from the line. For seven out of nine objects, the differences between \thb\
given by the two methods are less than 0.1 dex. The two exceptions are \iras\
and Mrk 1044. In Paper II we show that using the integration method, we cannot
obtain a significant \thb\ for \iras. The lag for Mrk 1044 reported in that
paper is about a factor 2 shorter than obtained here, with a very large
uncertainty ($4.8_{-3.7}^{+7.4}$). This value is consistent with the new one
presented here within the uncertainties. In addition, the uncertainty on \thb\
in Mrk 493 given by the integration method is very large with a negative lower
limit ($12.2_{-16.7}^{+3.5}$). The fitting method yields much smaller
uncertainty with basically the same \thb. This is a direct result of the much
smoother \hb\ light curve of Mrk 493 given by the fitting method (see Figure
\ref{fig-lc}, and also Figure 2 of Paper II). Thus, the time lags of \hb\
given by the two methods are consistent with each other and \thb\ is better
defined by the fitting procedure in some of the cases.

\section{Discussion}
\label{sec-discussion}

\subsection{The Radius--Luminosity Relationship for \feii}

A simple theoretical expectation based on photoionization is that $R_{\rm BLR}
\propto L^\alpha$ with $\alpha \sim 0.5$. The \rl\ relation for \hb\ has been
compared with such predictions in several earlier publications
\citep[e.g.,][and references therein]{kaspi00,bentz09,bentz13}. Regarding
\feii\ lines, \citet{chelouche14} provide a tentative \rl\ relation for the
first time, from a small inhomogeneous sample of six AGNs. Among them, two
objects are reported by \citet{barth13} using the standard spectral fitting
method, four others come from \citet{rafter13} and \citet{chelouche14} using
the MCF scheme of \citet{chelouche13}. Here, we revisit this issue by more
than doubling the size of the sample by adding our nine newly measured
sources. For a proper comparison with the \citet{chelouche14} results we used
our best observed fluxes and uncertainty and a cosmology with $H_0=70~{\rm
km~s^{-1}}$ Mpc$^{-1}$, $\Omega_{m}=0.3$, and $\Omega_\Lambda=0.7$.

\begin{figure}
  \centering
  \includegraphics[width=0.45\textwidth]{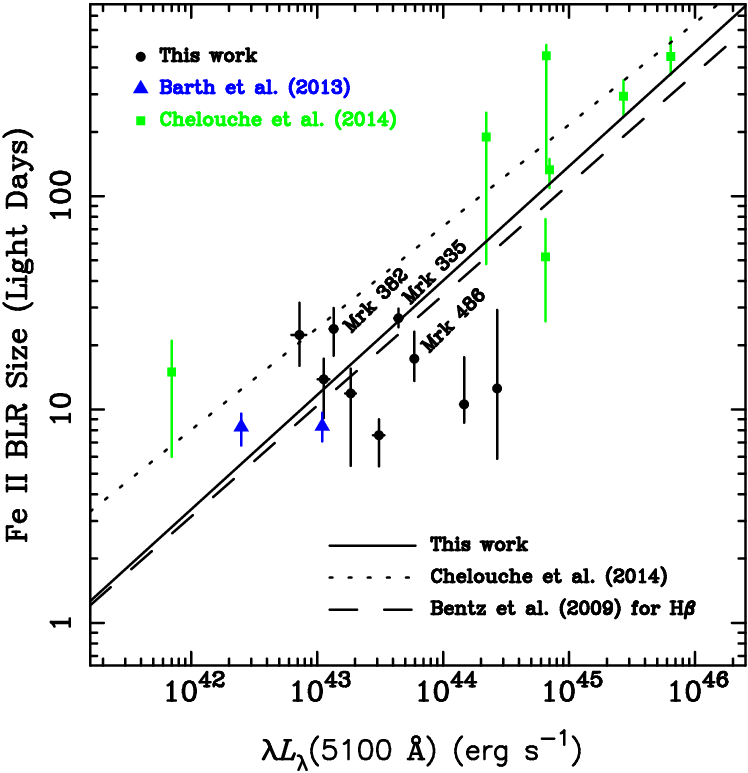}
  \caption{
  \feii\ \rl\ relation. The nine objects in our sample are plotted as black
  dots, the three objects with their names labeled have relatively large
  differences between \thb\ and \tfe. Measurements from previous studies are
  plotted as blue triangles and green squares (see text for details). The
  solid line is the fit to the entire data set. The dotted and dashed lines
  are the \rl\ relations in \citet{chelouche14} for \feii\ and \citet{bentz09}
  for \hb, respectively. 
  }
  \label{fig-ferl}
\end{figure}

Figure \ref{fig-ferl} shows \tfe\ vs. the 5100 \AA\ luminosity for the nine
objects in our sample (black dots) alongside the earlier results (kindly
provided by D. Chelouche). The green squares are those obtained by the MCF
scheme (including three with insignificant results, see Figure 4 of
\citealt{chelouche14} for details). The blue triangles are the two objects
from \citet{barth13} for which we used the centroid of the published CCFs for
\feii\ and the $V$-band light curves. Using the FITEXY method \citep{press92}
to fit all the 18 objects we find:
\begin{equation}
  {\rm log} (R_{{\rm Fe}~\textsc{ii}~{\rm BLR}}) = (-22.0 \pm 1.1) + 
  (0.54 \pm 0.02) {\rm log} (\lambda L_{\lambda} (5100~\mathring{\rm A}))~,
\end{equation}
which is plotted in a solid line. This line deviates considerably from the
relation presented in \citet{chelouche14} (shown in a dotted line), but close
to that of \citet{bentz09} for \hb\ (dashed line). This result is expected
since most of the objects in our sample show roughly equal \tfe\ and \thb\
(see Table \ref{tab-lag}), except for three objects (Mrk 335, Mrk382, and
Mrk486), while the previous studies have \tfe\ greater than \thb. Our sample
by itself shows no correlation at all between \tfe\ and the 5100 \AA\
luminosity. This is not surprising given the relatively small range in
luminosity considered by us and the fact that for this sample, the correlation
between \thb\ and the continuum luminosity is also very weak. We suggest that
this lack of correlation is related also to the large \er\ of the objects in
our sample, a topic which we discuss in great detail in Paper IV (Du et al.,
submitted).

\subsection{Comparison between \feii\ and \hb}
\label{sec-cmpfehb}

\begin{figure}
  \centering
  \includegraphics[width=0.45\textwidth]{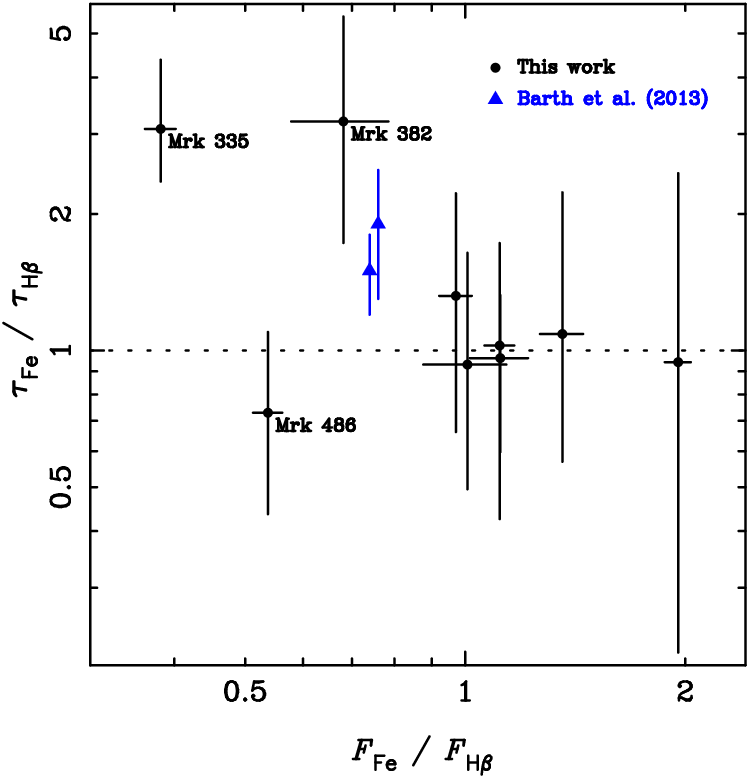}
  \caption{
  Time lag ratios of \feii\ to \hb\ vs. intensity ratios of \feii\ to \hb.
  Objects in our sample are shown as black dots. The three objects labeled by
  names have different time lags of \feii\ and \hb, and all have low intensity
  ratio of \feii\ to \hb. Blue triangles show the approximate positions of the
  two objects in \citet{barth13} (intensity ratios are estimated by eye). The
  two points are slightly shifted horizontally for clarity (See text for
  details.)
  }
  \label{fig-rfelag}
\end{figure}

As shown in Table \ref{tab-lag}, \tfe\ is roughly equal to \thb\ (counting the
uncertainties in both lags) in six of our objects, longer in two (Mrk 335 and
Mrk 382) and somewhat shorter in one (Mrk 486). A common feature of the three
objects with unequal lags is the relatively low \feii/\hb\ intensity ratio.
Figure \ref{fig-rfelag} shows the plot of \tfe/\thb\ vs. \ffe/\fhb\ (defined
as \rfe). The three objects with different time lags are labeled by their
names and locate in the region of \rfe\ $<$ 1. The six other objects show
\rfe\ $>$ 1. We also mark the approximate positions of the two objects
measured by \citet{barth13}.%
\footnote{
The \feii\ flux in \citet{barth13} is defined as the integrated flux of \feii\
between 4400 and 4900 \AA, while our definition refers to the range between
4434 and 4684 \AA. For the \feii\ template of \citet{boroson92}, \ffe\ defined
here is about 3/4 that in \citet{barth13}.
}
Based on eye estimates of the published light curves shown in their paper,
both objects have \rfe\ $\sim$ 0.75. Their positions in Figure
\ref{fig-rfelag} are marked around 0.75 (and horizontally shifted for
clarity). They both have different time lags than \hb\ and \rfe\ $<$ 1,
consistent with the results of our sample.

We note, again, that our sample is highly biased toward AGNs with high \er\
and thus high \rfe. In an unbiased AGN sample, the fraction of sources with
\rfe\ $<$ 1 is much larger (see, e.g., Figure 7 of \citealt{sulentic00} and
Figure 1 of \citealt{shen14}). \citet{hu08b} analysed the properties of a
large number of low-redshift AGNs. They found that the velocity width of
\feii\ is systematically narrower ($\sim 3/4$) than that of \hb, which may
suggest, given Keplerian velocities, that in those sources \tfe\ is longer
than \thb. Our present sample contains only 10 sources which is too few to
test, systematically, any of these ideas.

\section{Summary}
\label{sec-summary}

We provide new \feii\ measurements for 10 NLS1s and report statistically
significant time lag measurements for nine of these sources. All the observed
NLS1s are suspected to be SEAMBHs. This more than doubles the number of AGN
showing measurable \feii\ time lags. Our time lag measurements are based on
high-cadence, high-S/N measurements at the Yunnan observatory, and on a new,
sophisticated, fitting analysis that takes into account the uncertainties
caused by the apparent flux change of the host galaxy and several narrow
emission lines. We demonstrate, by a careful comparison with earlier
measurements of \thb, that the method can considerably improve the accuracy of
the time lag measurement. The main findings reported in this paper can be
summarized as follows:
\begin{enumerate}
\item
All 10 objects presented here show \feii\ variations with an amplitude of a
few to ten percent. On average, this is about 10\% smaller than the
variability of the \hb\ line.
\item
Reliable \feii\ reverberation time lags with respect to the AGN continuum are
detected in nine objects, confirming the suggestion that the \feii\ emission
originates from photoionized gas. 
\item
Combining the new reverberation mapping results with those in previous work,
shows a clear radius--luminosity relationship for \feii\ that is similar to
the one for \hb.  However, our sample by itself shows no such correlation due
to the large intrinsic scatter over a small luminosity range.
\item
The difference in the time lags of \feii\ and \hb\ depends on the intensity
ratio of \feii\ to \hb\ (\rfe). The time lag of \feii\ is roughly equal to
that of \hb\ in all the six objects with \rfe\ $\gtrsim$ 1. The \feii\ time
lag is longer in Mrk 335 and Mrk 382, those objects with \rfe\ $<$ 1, and
shorter in Mrk 486 with \rfe\ $<$ 1.
\end{enumerate}
 
\acknowledgments
We acknowledge the support of the staff of the Lijiang 2.4m telescope. Funding
for the telescope has been provided by CAS and the People's Government of
Yunnan Province. This research is supported by the Strategic Priority Research
Program -- The Emergence of Cosmological Structures of the Chinese Academy of
Sciences, grant No. XDB09000000, by the NSFC through NSFC-11173023, -11133006,
-11233003, -11473002, and by Israel-China ISF-NSFC grant 83/13. 

\appendix

\section{Host Galaxy Flux Calibration}
\label{sec-host}

Our flux calibration method provides the \textit{relative} flux of the object
with respect to that of the local comparison star. For two point sources kept
in a line parallel to the slit, the fractions of light loss due to seeing,
differential atmospheric refraction, and mis-centering are identical. Thus
this method gives accurate relative flux calibration for the spectral
components of the AGN, including the featureless power law (\fagn) and the
broad emission lines. This is not the case for the extended host galaxies
whose flux relative to the comparison star inside the slit can vary much more
due to seeing variations and mis-centering. In addition, different parts of
the galaxy may be observed each time. As a result, the derived flux of the
host galaxy is $F_{\rm gal} = f_{\rm cal} F_{\rm gal,abs}$, where $F_{\rm
gal,abs}$ is the absolute flux (which is constant), and $f_{\rm cal}$ is a
factor accounting for all the effects of varying observing conditions in the
flux calibration procedure. $f_{\rm cal}$ could change from one exposure to
the next, introducing an additional uncertainty into the integrated 5100 \AA\
flux $F_{\rm 5100} = F_{\rm AGN} + f_{\rm cal} F_{\rm gal,abs}$.

In Papers I and II, we estimated the flux of the host galaxies in the spectral
extraction aperture using archival \textit{HST} images for eight objects in
our sample. The resultant relative fluxes, (\fgal/\fagn), range from $\sim$0.2
in Mrk 335 to $\gtrsim$ 1 in Mrk 382. Thus, the apparent change of flux of the
host galaxy is not negligible for those objects with strong host galaxy
contribution. For Mrk 382, the measurement of the AGN continuum was badly
affected by this uncertainty, which forced us to use the $V$-band photometry
instead. The new procedure adopted here (see the main text) enables us to
solve for $f_{\rm cal} F_{\rm gal,abs}$ for each individual-night spectrum.
The detailed fitting of the various spectral components considerably reduces
the noise in the various line and continuum light curves. In particular, for
Mrk 382, we can now recover the 5100 \AA\ AGN continuum variations to a much
better accuracy, which is evident by the good agreement with the $V$-band
photometry.

\begin{figure}
  \centering
  \includegraphics[width=0.4\textwidth]{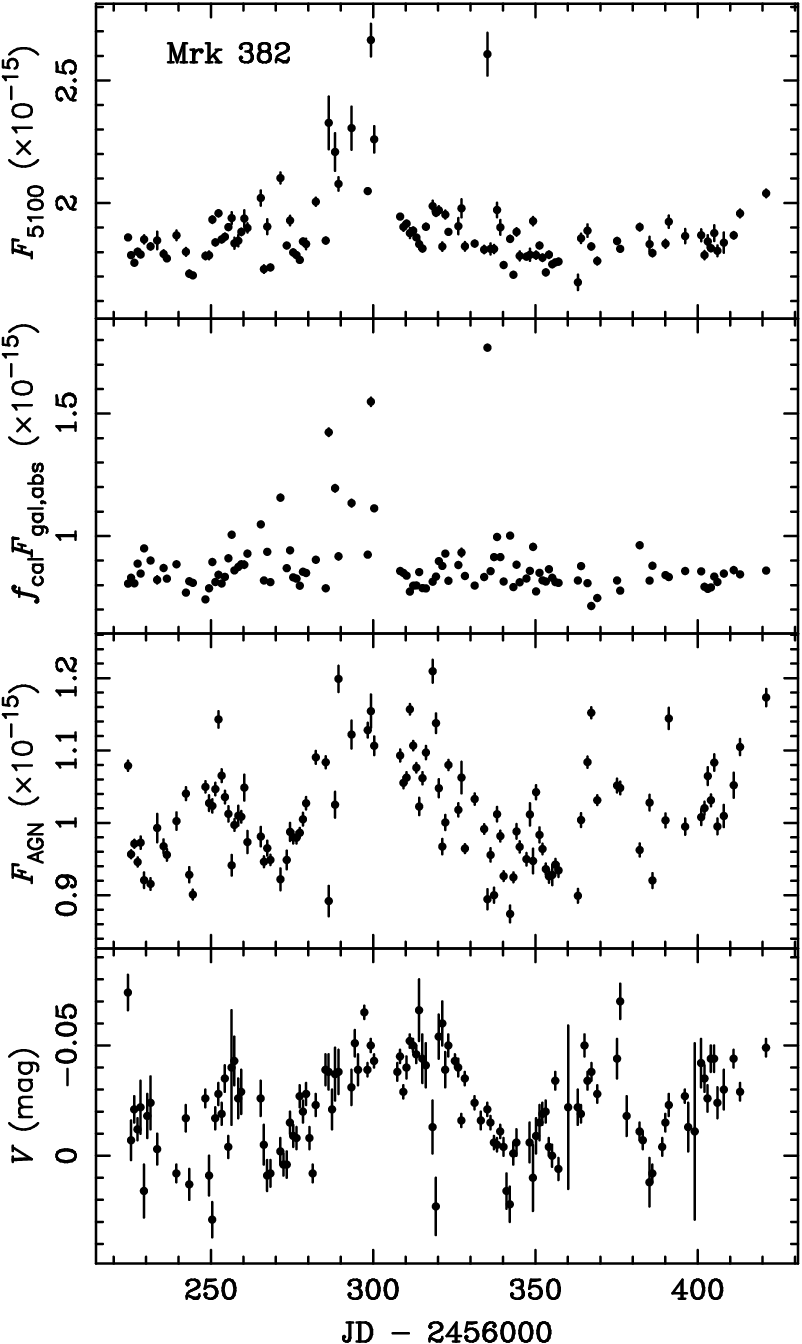}
  \caption{
  Light curves of the integrated 5100 \AA\ flux (\fsumc, top panel), fitted
  host galaxy ($f_{\rm cal} F_{\rm gal,abs}$ described in Appendix
  \ref{sec-host}, upper-middle panel), fitted AGN power-law continuum (\fagn,
  lower-middle panel), and $V$-band magnitude (bottom panel, see also Paper
  II). Note that \fsumc\ is heavily contaminated by the host whose apparent
  flux is varying by $f_{\rm cal}$, and the \fagn\ recovers, accurately, the
  intrinsic AGN variability as judged by a comparison with the $V$-band
  observations.
  }
  \label{fig-lc382host}
\end{figure}

The big improvements due to the spectral fitting procedure are shown in Figure
\ref{fig-lc382host}, which provides more information on the process for Mrk
382. The top panel shows the light curve of \fsumc, which is measured as in
Papers I and II, except for the inclusion of Galactic extinction correction
and de-redshifting. The scatter in this light curve is very large, and hence
it was replaced in Paper II by the $V$-band light curve. The upper-middle and
lower-middle panels show the results of the spectral fitting for \fgal\ and
\fagn, respectively. The light curve of \fagn\ is basically identical to that
of the $V$-band (bottom panel). Clearly, the unrealistic high \fsumc\ around
JD 2456300 was caused by the large contribution of the host galaxy caused by
seeing and guiding fluctuations. On other JDs, the host contamination is
relatively small, but still strong enough to affect the variability of \fagn,
e.g., the peak around JD 2456250.

\section{\feii\ Light Curves Measured by Integration}
\label{sec-feintg}

\begin{figure*}
  \centering
  \includegraphics[width=0.8\textwidth]{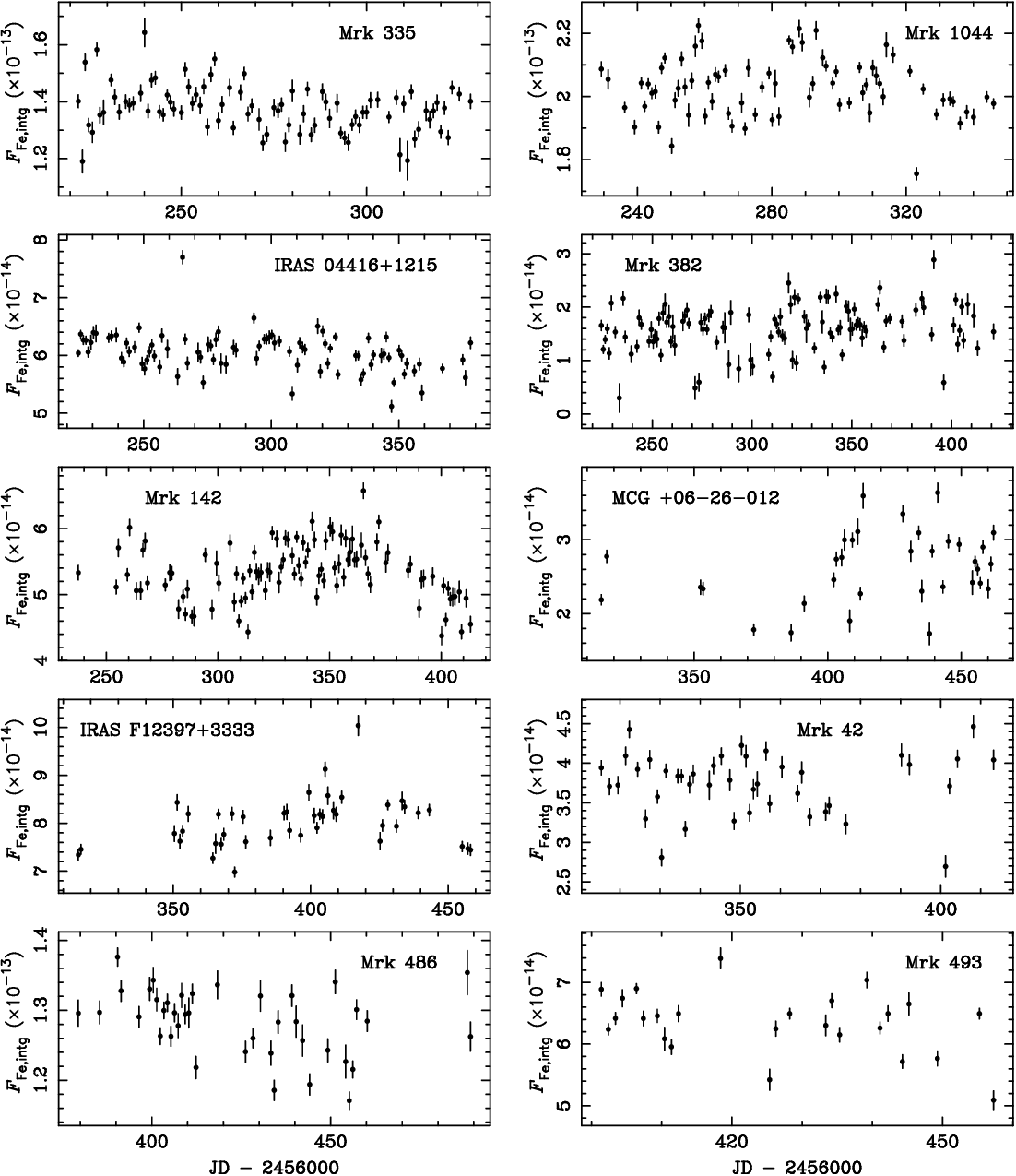}
  \caption{
  Light curves of \feii\ measured by simple integration ($F_{\rm Fe,intg}$).
  All have larger scatter comparing with those obtained by fitting shown in
  Figures \ref{fig-lc} and \ref{fig-lc42}.
  }
  \label{fig-feintg}
\end{figure*}

We present the \feii\ light curves measured by the traditional integration
method for a comparison with that by the present fitting scheme in this
Appendix. The flux of the \feii\ bump on the red side of \oiii\ $\lambda$5007
is calculated by integrating in 5115--5465 \AA\ the flux above the straight
line set by two continuum windows 5085--5115 and 5465--5495 \AA\ (the three
bands are in the rest frame). Figure \ref{fig-feintg} shows these light
curves. Comparing with those given by the fitting method (in Figures
\ref{fig-lc} and \ref{fig-lc42}), all the light curves by integration show
larger scatter. For four objects (Mrk 1044, Mrk 142, \mcg, and \irasf), rough
structures can be recognized in their light curves, and the derived \feii\
time lags are consistent with those given by the fitting but with larger
uncertainties. The remaining objects fail to have reliable measurement of time
lag, for their scattered light curves without clear structures. We also tried
to measure the flux of the \feii\ bump on the blue side of \hb\ by choosing an
integration interval avoiding the broad \heii\ line. Similarly, light curves
with poor quality are obtained.

\end{document}